\documentclass[]{elsart}

\usepackage{graphicx} \usepackage{dcolumn} \usepackage{bm}
\usepackage{subfigure} \usepackage{epsfig} \usepackage{natbib}
\usepackage{graphicx} \usepackage{latexsym} \usepackage{amsmath}
\usepackage{rotating} \rotdriver{dvips}

\newcommand{\vect}[1]{\mbox{\boldmath$#1$\unboldmath}}
\newcommand{\bigO}[1]{{O}(#1)}

\newcommand{\CK}{CK}
\newcommand{\degc}{$^\circ$C}
\newcommand{\ie}{{\it i.e.}}
\newcommand{\eg}{{\it e.g.}}

\newcommand{\bsl}{{\ell}}

\newcommand{\Rey}{\mbox{Re}}

\newcommand{\kb}{k}
\newcommand{\ka}{\kappa a}
\newcommand{\kL}{\kappa L}

\newcommand{\grad}{\vect{\nabla}}
\newcommand{\lapl}{\nabla^2}
\newcommand{\dive}{\vect{\nabla} \cdot}

\begin{document}

\begin{frontmatter}


\title{`Exact' solutions of the full electrokinetic model for soft
spherical colloids: Electrophoretic mobility}

\author{Reghan J. Hill\corauthref{cor1}} \ead{reghan.hill@mcgill.ca}
\corauth[cor1]{Corresponding author} \address{Department of Chemical
Engineering and McGill Institute for Advanced Materials, McGill
University, Montreal, Quebec, H3A2B2, CANADA}

\author{D. A. Saville} \address{Department of Chemical Engineering,
Princeton University, Princeton, New Jersey, 08542, USA}

\begin{abstract}
  Numerical solutions of the standard electrokinetic model provide a
  basis for interpreting a variety of electrokinetic phenomena
  involving `bare' colloids. However, the model rests on the classical
  notion of a shear or slipping plane, whose location is unknown when
  surfaces are coated with permeable polymer. Consequently, an
  electrokinetic model for `soft', `hairy' or `fuzzy' colloids has
  been developed, but until recently solutions were available only for
  several restricted cases, most notably for particles with thin,
  uniform layers, and without polarization and relaxation. Here we
  present numerically exact solutions of the full model for a variety
  of soft colloids, including PEG-coated liposomes, PEO-coated
  latices, human erythrocytes, and polyelectrolyte
  micelles. Particular attention is given to linking the thickness,
  density and permeability of the coatings, which are key parameters
  in the model, to ``physical'' quantities, such as the polymer
  molecular weight, adsorbed amount, and hydrodynamic layer
  thickness. This paper also identifies limits---on the ionic
  strength, particle size, layer thickness and permeability---beyond
  which earlier theories breakdown. In short, polarization and
  relaxation are as influential on the mobility of soft colloids as
  they are for `bare' particles.
\end{abstract}

\end{frontmatter}

\tableofcontents

\section{Introduction} \label{sec:introduction}

Polymer deposited on the surface of `bare' colloidal particles can be
used to regulate the stability and rheology of dispersions; the
influence of {\em thin} polymer coatings on the drag coefficient and
intrinsic viscosity, for example, are available from exact solutions
of the Stokes equations~\citep[]{Anderson:1996}. In addition,
adjusting the electrostatic, steric, and dispersive contributions to
the inter-particle potential affords control over the kinetics and
thermodynamics of self-assembly processes. In turn, these bear
directly on the microstructure of materials with novel
properties. Nevertheless, the polymer coating density, thickness and
charge affect electrokinetic transport processes near the supporting
surface, making it difficult to discern competing influences that
affect light scattering and electrophoresis measurements, for
example. Clearly, the application of polymer coatings for tailoring
the properties of dispersions will benefit from a reliable
electrokinetic theory.

The purpose of this paper is to describe applications of a recently
developed model for predicting the influence of polymer coatings on a
variety of electrokinetic particle characterization measurements. The
focus of this paper is on the electrophoretic mobility, but it should
be kept in mind that the model permits solutions under dynamic
conditions, as encountered in dielectric relaxation spectroscopy and
electro-sonic-amplitude devices.

While the electrokinetics of bare particles are (relatively) well
understood, the situation for colloids coated with a permeable polymer
is less established. With bare particles, ion transport and
hydrodynamics follow classical models, where, for example,
hydrodynamic shear begins at the particle surface (shear
surface). Efforts to model transport processes inside a thin porous
layer have adopted a variety of
approaches~\citep{Dukhin:1970,Rosen:1993}. More realistic (and
physically appealing) models treat polymer layers as continuous porous
media with a permeability that reflects the density and hydrodynamic
size of the polymer segments.

Charged layers, which are important because of their biological
context, present a challenging set of problems. A variety of
approximations have been used to construct tractable models, but,
until recently, none had the generality of O'Brien and White's
numerical solution of the standard model for bare
particles~\citep{OBrien:1978}. For example, Brooks and coworkers
developed models for human erythrocytes (red blood cells), with
`flat', uniformly charged layers to mimic the extracellular glycocaylx
layer~\citep{Levine:1983,Sharp:1985}; Hermans and Fujita considered
porous charged spheres~\citep{Hermans:1955b,Hermans:1955a}; and, more
recently, Ohshima and coworkers generalized these models, presenting
approximate solutions (for the electrophoretic mobility) that neglect,
in particular, polarization and
relaxation~\citep{Ohshima:1989,Ho:1996}.

Without exact solutions, the accuracy of approximations is unknown, so
empirical parameters derived from experiments using approximate
theories should be accepted with caution. Saville demonstrated that
even `large' particles with thin polymer and diffuse double layers are
susceptible to polarization and
relaxation~\citep{Saville:2000}. Inferences drawn from his
semi-analytical theory have been borne by numerically exact solutions
of the full model~\citep{Hill:2003a}, and recent attention has turned
to details of the polymer segment density
distribution~\citep{Hill:2004a}.

This paper reviews the full electrokinetic model for colloids with
soft neutral and charged layers. We draw upon numerically exact
solutions of the governing equations, as provided by Hill, Saville and
Russel's model~\citep{Hill:2003a}. Their methodology removes all
restrictions imposed by earlier approximate theories, permitting
studies that examine the influence of particle size and polymer-layer
structure in considerable detail. Accordingly, this paper establishes
the parameter spaces of earlier approximate theories as limiting cases
of the full model.

We begin with a presentation of the full electrokinetic
model~(\S~\ref{sec:theory}), linking the solution of the equations to
the mobility and effective layer thickness~(\S~\ref{sec:mobility}),
and briefly discussing key parameters that characterize polymer
layers~(\S~\ref{sec:polymerlayer}). Next, the results are presented,
beginning with particles that have {\em neutral} polymer
coatings~(\S~\ref{sec:neutral}). First, we highlight the significant
role of the hydrodynamic layer thickness, distinguishing it from the
actual thickness~(\S\ref{sec:highionicstrength}). Detailed comparisons
are made with Ohshima's theories for thin coatings at high ionic
strength, and, in particular, establishing the significance of
polarization and relaxation~(\S~\ref{sec:polarization}). Next, a
methodology for interpreting experimental data is
presented~(\S~\ref{sec:interpretation}), and the full model is then
applied to interpret the mobilities of stealth
liposomes~(\S~\ref{sec:peg}) and PEO-coated
latices~(\S~\ref{sec:peo}). Turning to particles with {\em charged}
layers~(\S~\ref{sec:charged}), we compare the full model with
approximate theories for thin layers~(\S~\ref{sec:thinpoly}). `Exact'
calculations are presented for human
erythrocytes~(\S~\ref{sec:bloodcells}) and `small', highly charged
polyelectrolyte micelles~(\S~\ref{sec:diblocks}). We conclude with
summary~(\S~\ref{sec:summary}).

\section{Theory} \label{sec:theory}

The electrokinetic model and the computational methodology adopted in
this work have been described in detail by Hill, Saville and
Russel~\citep{Hill:2003a}. The model augments the standard
electrokinetic model for bare colloids~\citep{OBrien:1978} with
arbitrary (radial) distributions of Stokes resistance centers and
charge. Electroosmotic flow within the polymer layer is hindered by
hydrodynamic drag on the segments. In this work, electromigration and
molecular diffusion are unaltered, which is reasonable when ($i$) the
hydrodynamic volume fraction of the ions is low and ($ii$) the ions
are small compared to the polymer intersties, \ie, the polymer volume
fraction is low. In principle, however, the methodology permits the
introduction of spatially dependent ion mobilities, solvent viscosity
and dielectric constant.

\subsection{The electrokinetic transport equations}

The transport equations and boundary conditions are presented here in
dimensional form. As usual, they comprise the non-linear
Poisson-Boltzmann equation
\begin{equation} \label{eqn:pbeqn}
  \epsilon_o \epsilon_s \lapl \psi = - \sum_{j=1}^{N} (n_j - n^f_j)
  z_j e,
\end{equation}
where $\epsilon_o$ and $\epsilon_s$ are the permittivity of a vacuum
and dielectric constant of the electrolyte (solvent); $n_j$ and
$n^f_j$ are the concentrations of the $j$th mobile and fixed charges,
with $z_j$ the valences\footnote{The valence of the fixed charge is
set opposite to that of its respective (mobile) counterion in
Eqn.~(\ref{eqn:pbeqn})}; $e$ is the elementary charge and $\psi$ the
electrostatic potential.

Transport of the mobile ions is governed by
\begin{equation} \label{eqn:iontransport}
  6 \pi \eta a_j (\vect{u} - \vect{v}_j) - z_j e \grad \psi - \kb T
  \grad \ln{n_j} = 0 \ \ (j=1,...,N),
\end{equation}
where $a_j$ are Stokes radii, obtained from limiting conductances or
diffusivities; $\eta$ is the electrolyte viscosity, and $\vect{u}$ and
$\vect{v}_j$ are the fluid and ion velocities; $\kb T$ is the thermal
energy.

Ion conservation demands
\begin{equation} \label{eqn:ionconservation}
  \partial n_j / \partial t = -\dive (n_j \vect{v}_j) \ \ (j=1,...,N),
\end{equation}
where $t$ denotes time, with the ion fluxes $\vect{j}_j = n_j
\vect{v}_j$ obtained from Eqn.~(\ref{eqn:iontransport}).

Similarly, momentum and mass conservation require
\begin{equation} \label{eqn:linnseqns}
  \rho_s \partial \vect{u} / \partial t = \eta \lapl \vect{u} - \grad
  p + \eta / \bsl^2 (\vect{V} - \vect{u}) - \sum_{j=1}^{N} n_j z_j e
  \grad \psi
\end{equation}
and
\begin{equation} \label{eqn:incomp}
  \dive \vect{u} = 0,
\end{equation}
where $\rho_s$ and $\vect{u}$ are the electrolyte (solvent) density
and velocity, and $p$ is the pressure. Note that $\eta / \bsl^2
(\vect{V} - \vect{u})$ represents the hydrodynamic drag force exerted
by the polymer on the interstitial fluid, with $\bsl^2$ the (radially
varying) permeability of the polymer, and $\vect{V}$ is the particle
velocity. Clearly, the polymer and particle are assumed to move as a
rigid composite, and the particle Reynolds number $\Rey = V a \rho_s /
\eta$ is assumed small.

As usual, the double-layer thickness
\begin{equation} \label{eqn:kappa}
  \kappa^{-1} = [ \kb T \epsilon_s \epsilon_o / (2 I e^2)]^{1/2}
\end{equation}
emerges from Eqns.~(\ref{eqn:pbeqn}) and~(\ref{eqn:iontransport})
where
\begin{equation} \label{eqn:ionicstrength}
  I = (1/2) \sum_{j=1}^{N} z^2_j n^\infty_j
\end{equation}
is the bulk ionic strength, with $n_j^\infty$ the bulk ion
concentrations. Note that ion diffusion coefficients are
\begin{equation} \label{eqn:diffusivity}
  D_j = \kb T / (6 \pi \eta a_j),
\end{equation}
and the permeability (square of Brinkman screening length) may be
expressed as
\begin{equation} \label{eqn:brinkmanscreeninglength}
  \bsl^2 = 1 / (n 6 \pi a_s F_s) = 2 a_s^2 / (9 \phi_s F_s),
\end{equation}
where $n(r)$ is the concentration of Stokes resistance centers
(segments), with $a_s$ and $F_s(\phi_s)$ the Stokes radius and drag
coefficient of the segments.  For the rigid, `random' configurations
of spherical segments assumed in this work, the segment drag
coefficient varies with the hydrodynamic volume fraction $\phi_s = n
(4/3) \pi a_s^3$ according to~\citep{Koch:1999}
\begin{equation} \label{eqn:brinkmandragcoeff}
  F_s = \frac{1 + 3 (\phi_s/2)^{1/2} + (135/64) \phi_s \ln{\phi_s} +
    16.456 \phi_s}{1 + 0.681 \phi_s - 8.48 \phi_s^2 + 8.16 \phi_s^3} \ \
  (\phi_s < 0.4).
\end{equation}
The numerator in Eqn.~(\ref{eqn:brinkmandragcoeff}) derives from
theory for low to moderate volume fractions~\citep{Kim:1985}, with the
denominator providing a smooth transition through data from multipole
simulations ($\phi_s < 0.4$) to the well-known Carman correlation
($0.4 < \phi_s < 0.64$). Note that other drag-coefficient correlations
exist for {\em porous media} with ordered and other random
microstructures~\citep[see][for fibrous
microstructures]{Howells:1998,Jackson:1986}. At present, however, such
details are of secondary importance to the distribution of segments
$n(r)$ and their size $a_s$. The segment and fixed-charge density
distributions are discussed in section~\ref{sec:polymerlayer}, and
throughout sections~\ref{sec:neutral} and~\ref{sec:charged}.

\subsection{Inner (particle surface) boundary conditions}

The model permits either the equilibrium surface potential $\zeta$ or
surface charge density $\sigma_c$ to be specified. Because the surface
($r = a$) is assumed impermeable, the surface charge is constant under
the influence of an external electric field or particle motion. This
is achieved with no-flux boundary conditions for each (mobile) ion
species. As usual, the no-slip boundary condition applies to the
electrolyte. It follows that (inner) boundary conditions are either
\begin{equation} \label{eqn:zetapotential}
  \psi = \zeta \mbox{ at } r = a
\end{equation}
or
\begin{equation} \label{eqn:psibc1}
  \epsilon_s \epsilon_o \grad \psi |_{out} \cdot \hat{\vect{n}} -
  \epsilon_p \epsilon_o \grad \psi |_{in} \cdot \hat{\vect{n}} = -
  \sigma_c \mbox{ at } r = a,
\end{equation}
with
\begin{equation} \label{eqn:impenetrable}
  n_j \vect{v}_j \cdot \hat{\vect{n}} = 0 \mbox{ at } r = a \ \ (j=1,...,N)
\end{equation}
 and
\begin{equation} \label{eqn:noslip}
  \vect{u} = \vect{V} \mbox{ at } r = a.
\end{equation}
Note that $\hat{\vect{n}} = \vect{e}_r$ is an outward unit normal and
$\vect{r} = r \vect{e}_r$ is the (radial) distance from the center of
the particle. The dielectric constant of the bare particle
$\epsilon_p$ does not affect the electrophoretic mobility to linear
order in the applied electric field~\citep{OBrien:1978}, but it is
relevant under dynamic (high-frequency) forcing, such as in dielectric
spectroscopy.

\subsection{Outer (far-field) boundary conditions}

Particle interactions are neglected, so an otherwise stationary,
infinite domain leads to far-field boundary conditions
\begin{equation} \label{eqn:psibc2}
  \psi \rightarrow - \vect{E} \cdot \vect{r} \mbox{ as } r \rightarrow
  \infty,
\end{equation}
\begin{equation} \label{eqn:bulkconc}
  n_j \rightarrow n_j^\infty \mbox{ as } r \rightarrow \infty \ \ (j=1,...,N),
\end{equation}
and
\begin{equation} \label{eqn:rest}
  \vect{u} \rightarrow 0 \mbox{ as } r \rightarrow \infty,
\end{equation}
where $\vect{E}$ is a uniform applied electric field.

\subsection{Solution of the equations}

Coupling the transport equations and boundary conditions to the
particle equation of motion specifies the electrokinetic
model\footnote{A computer package (MPEK) implementing Hill, Saville
and Russel's methodology is currently available from the corresponding
author.}. As usual, the solution requires non-linear perturbations to
the equilibrium state to be neglected. Perturbations (proportional to
the applied electric field) are introduced via
\begin{equation}
  \psi = \psi^0 - \vect{E} \cdot \vect{r} + \psi',
\end{equation}
\begin{equation}
  n_j = n_j^0 + n'_j \ \ (j=1,...,N) ,
\end{equation}
and
\begin{equation}
  p = p^0 + p',
\end{equation}
where the first terms on the right-hand sides denote the equilibrium
values, and the primed quantities denote the perturbations.

\subsection{Equilibrium state}

When $\vect{E} = \vect{V} = 0$, equilibrium is specified according to
\begin{equation}
  \epsilon_o \epsilon_s \lapl \psi^0 = - \sum_{j=1}^{N} (n^0_j -
  n^f_j) z_j e,
\end{equation}
\begin{equation}
  0 = - \dive{[- D_j \grad{n^0_j} - z_j e D_j / (\kb T) n^0_j \grad{\psi^0}]} \ \ (j=1,...,N),
\end{equation}
and
\begin{equation}
  0 = - \grad p^0 - \sum_{j=1}^{N} n^0_j z_j e \grad \psi^0,
\end{equation}
with either
\begin{equation}
  \psi^0 = \zeta
\end{equation}
or
\begin{equation}
  \epsilon_s \epsilon_o \grad \psi^0 |_{out} \cdot \vect{e}_r -
  \epsilon_p \epsilon_o \grad \psi^0 |_{in} \cdot \vect{e}_r =
  - \sigma_c \mbox{ at } r = a,
\end{equation}
\begin{equation}
  [- D_j \grad{n^0_j} - z_j e D_j / (\kb T) n^0_j \grad{\psi^0}] \cdot
  \vect{e}_r = 0 \mbox{ at } r = a \ \ (j=1,...,N),
\end{equation}
and
\begin{equation}
  \psi^0 \rightarrow 0 \mbox{ as } r \rightarrow \infty,
\end{equation}
\begin{equation}
  n^0_j \rightarrow n_j^\infty \mbox{ as } r \rightarrow \infty \ \ (j=1,...,N).
\end{equation}

\subsection{Linearized perturbed state}

Under periodic forcing
\begin{equation}
  \vect{E} = E \exp{(-i \omega t)}\vect{e}_z,
\end{equation}
with frequency $\omega / (2 \pi)$, axisymmetric perturbations take the
forms
\begin{equation}
  \psi' = \hat{\psi}(r) \vect{E} \cdot \vect{e}_r,
\end{equation}
\begin{equation}
  n'_j = \hat{n}_j(r) \vect{E} \cdot \vect{e}_r \ \ (j=1,...,N)
\end{equation}
and
\begin{equation}
  \vect{u} = - 2 (h_r / r) (\vect{E} \cdot \vect{e}_r) \vect{e}_r -
  (h_{rr} + h_{r} / r) (\vect{E} \cdot \vect{e}_\theta)
  \vect{e}_\theta,
\end{equation}
where
\begin{equation}\label{eqn:velocity}
  \vect{u} = \grad \times \grad \times h(r) \vect{E}.
\end{equation}
The perturbations satisfy
\begin{equation}
  \epsilon_o \epsilon_s \lapl \psi' = - \sum_{j=1}^{N} n'_j z_j e,
\end{equation}
\begin{equation}
  \partial n'_j / \partial t = - \dive{\vect{j}'_j} \ \ (j=1,...,N),
\end{equation}
where
\begin{eqnarray}
  \vect{j}'_j = - D_j \grad{n'_j} - z_j e D_j / (\kb T) n'_j
  \grad{\psi^0} - z_j e D_j / (\kb T) n^0_j (\grad{\psi'}-\vect{E}) \\
  + n^0_j \vect{u}, \nonumber
\end{eqnarray}
\begin{eqnarray}
  \rho_s \partial \vect{u} / \partial t = \eta \lapl \vect{u} - \grad
  p' - \eta / \bsl^2 (\vect{u} - \vect{V}) - \sum_{j=1}^{N} n^0_j z_j
  e (\grad \psi' - \vect{E}) \\ - \sum_{j=1}^{N} n'_j z_j e \grad \psi^0 \nonumber
\end{eqnarray}
and
\begin{equation}
  \dive \vect{u} = 0,
\end{equation}
with
\begin{equation}
  \epsilon_s \epsilon_o (\grad \psi' - \vect{E}) |_{out} \cdot
  \vect{e}_r - \epsilon_p \epsilon_o (\grad \psi' - \vect{E})
  |_{in} \cdot \vect{e}_r = 0 \mbox{ at } r = a,
\end{equation}
\begin{equation}
  \vect{j}_j' \cdot \vect{e}_r = 0 \mbox{ at } r = a \ \ (j=1,...,N),
\end{equation}
\begin{equation}
  \vect{u} = \vect{V} \mbox{ at } r = a,
\end{equation}
and
\begin{equation}
  \psi' \rightarrow 0 \mbox{ as } r \rightarrow \infty,
\end{equation}
\begin{equation}
  n'_j \rightarrow 0 \mbox{ as } r \rightarrow \infty \ \ (j=1,...,N),
\end{equation}
\begin{equation} 
  \vect{u} \rightarrow 0 \mbox{ as } r \rightarrow \infty.
\end{equation}

\subsection{Measurable quantities derived from the solution} \label{sec:mobility}

The linearized equations and boundary conditions, which are
characterized by multiple and widely varying length scales, are solved
using the methodology described in detail by Hill, Saville and
Russel~\citep[see][]{Hill:2003a}. In practice, the equations are
solved with the particle fixed at $\vect{r} = \vect{0}$, with ($i$) a
prescribed far-field velocity $-\vect{V}$ and $\vect{E} =
\vect{0}$---the so-called (U) problem, or ($ii$) a prescribed electric
field $\vect{E}$ and stationary far-field---the so-called (E)
problem. These solutions are superposed to satisfy the particle
equation of motion~\citep[see][]{OBrien:1978,Hill:2003a}, which leads
to the {\em electrophoretic} mobility,
\begin{equation} \label{eqn:stmobility}
  V / E = C^E / C^U,
\end{equation}
where
\begin{equation}
  C^X = \lim_{r \rightarrow \infty} h^X_r.
\end{equation}
Under oscillatory conditions, the {\it dynamic mobility} is
\begin{equation} \label{eqn:osmobility}
  V / E = C^E / [C^U - (\ka)^3 (\rho_p - \rho_s) / (3 \rho_s)],
\end{equation}
where
\begin{equation}
  C^X = \lim_{r \rightarrow \infty} h^X_r r^2
\end{equation}
and $\rho_p$ is the density of the bare particle. Note that the
superscripts $X \in \{U, E \}$ denote the (U) and (E) problems, where
$C^X$ are ``asymptotic coefficients'' that characterize the far-field
decay of $h(r)$ in Eqn.~(\ref{eqn:velocity}).

Following standard convention, the scaled {\em electrophoretic
  mobility} is defined as
\begin{equation} \label{eqn:mobility}
  M = (V / E) 3 \eta e / (2 \epsilon_s \epsilon_o \kb T),
\end{equation}
so the well-known H{\"u}ckel and Smoluchowski mobilities are $M =
\zeta$ and $M = (3/2) \zeta$, where $\zeta$ is scaled with $\kb T /
e$.

The {\em drag coefficient}
\begin{equation} \label{eqn:dragcoeff}
  F = f / (6 \pi \eta a V)
\end{equation}
is the force $f$ required to translate the particle with velocity $V$
in the absence of an applied electric field, scaled with the Stokes
drag force on the bare uncharged sphere. Frictional drag on the
polymer coating and electro-viscous effects arising from charge
increase the effective particle size. This motivates our definition of
an {\em effective coating thickness}
\begin{equation} \label{eqn:effthickness}
  L_e = a (F - 1),
\end{equation}
which, in the absence of electro-viscous stresses, is termed the {\em
hydrodynamic coating thickness}
\begin{equation} \label{eqn:hydrodynamicthickness}
  L_h = L_e \mbox{ when } \sigma_c = n^f_j = 0.
\end{equation}
In practice, the hydrodynamic coating thickness is realized when the
surface potential is low, which is often the case at high ionic
strength.

\subsection{Polymer layer characterization} \label{sec:polymerlayer}

The length and molecular weight of a polymer segment, $l$ and $M_s$,
and the number of segments per chain $N$ are defined so the chain
contour length and molecular weight are $l_c = N l$ and $W = N
M_s$. If the segments are monomer units, then $l$ and $N$ are denoted
$l_m$ and $N_m$, so $l_c = N_m l_m$ and $W = N_m M_m$. Similarly, if
the segments are defined as statistical or Kuhn segments, then the
mean-squared end-to-end distance of a free chain in a theta solvent is
$R = l N^{1/2} = l [(M/M_m)(l_m/l)]^{1/2}$, giving
\begin{equation}
  l = R^2 M_m / (M l_m).
\end{equation}
This is useful because $R$ can be expressed in terms of the
hydrodynamic radius and radius of gyration of `free' polymer chains,
and $l_m$ is usually known from X-ray crystallography or knowledge of
bond lengths and monomer architecture.

The mass of polymer per unit area of substrate is 
\begin{equation}
  W \sigma_p = N M_s \sigma_p = N_m M_m \sigma_p,
\end{equation}
where $\sigma_p$ is the number of chains per unit of surface area. The
number of segments per unit volume $n$ is often scaled to give a
volume fraction $\phi = n l^3$. The `actual' volumes of a solvent
molecule and a polymer segment are $v_1$ and $v_2$, so the physical
volume fraction of polymer is $n v_2$. Similarly, the hydrodynamic
volume fraction of polymer segments is $\phi_s = n (4/3) \pi a_s^3$,
where $a_s$ is the hydrodynamic (Stokes) radius of a segment.

The grafting density is often scaled to give $\sigma_p l^2$, but a
better qualitative `measure' of the grafting density is, perhaps,
$\sigma_p R^2 = \sigma_p N l^2 = \sigma_p (W/M_m) l_m l$. When
$\sigma_p N l^2 \sim 1$, the layer is considered semi-dilute ($\phi
\ll 1$) and, therefore, laterally homogeneous. When $\sigma_p N l^2 <
1$, however, the layer is dilute and, therefore, laterally
inhomogeneous. Moreover, when $\sigma_p N l^2 > 1$, layers are
homogeneous throughout and are referred to as brush-like. This
`picture' applies to di-block copolymers where one (shorter) block
anchors the other (longer) block to the substrate. The segments of
adsorbed homopolymers interact strongly with the surface, giving rise
to a dense inner region with a sparse periphery. The structure of
polyelectrolytes is more complicated, since the statistical segment
length varies with the bulk ionic strength and the degree of
counterion dissociation, the latter of which may depend on the ionic
strength and pH~\citep{Biesheuvel:2004}.

\section{The mobility of colloids with neutral coatings} \label{sec:neutral}

\subsection{Connection of the electrophoretic mobility to the hydrodynamic layer thickness} \label{sec:thinneutralcoatings}

In contrast to approximate analytical solutions, numerically exact
solutions of the full electrokinetic model permit arbitrary segment
and charge density distributions. Here we compare the electrophoretic
mobilities of particles with uniform and non-uniform (Gaussian-like)
layers, each with the same adsorbed amount and the same hydrodynamic
layer thickness.  In turn, the hydrodynamic layer thickness emerges as
a characteristic of prime importance.

Gaussian-like distributions
\begin{equation} \label{eqn:gaussiana}
  n(r) = n_1 \exp{[-(r - a)^2/\delta^2]} \ \ (r>a),
\end{equation}
where $n_1$ specifies the density at the bare surface ($r = a$) and
$\delta$ characterizes the layer thickness, mimic terminally anchored
polymer in a theta solvent when the molecular weight and grafting
density are moderately low. Substituting Eqn.~(\ref{eqn:gaussiana})
into the general relationship
\begin{equation} \label{eqn:densityconstraint}
  N \sigma_p a^2 = \int_a^\infty n(r) r^2 \mbox{d}r,
\end{equation}
where $\sigma_p$ is the chain grafting density and $N$ is the number
of segments per chain, gives
\begin{equation} \label{eqn:gaussianb}
  N \sigma_p a^2 = (n_1 \delta^3 / 4) \{\sqrt{\pi} [2 (a / \delta)^2 +
    1] + 4 a / \delta\}.
\end{equation}
Similarly, for a uniform distribution with density $n_0$ and thickness
$L$,
\begin{equation} \label{eqn:uniform}
  N \sigma_p a^2 = (n_0 / 3) a^3 [(1 + L / a)^3 - 1].
\end{equation}

\begin{figure}
\centering
\vspace{1cm}
  \includegraphics[height=7cm]{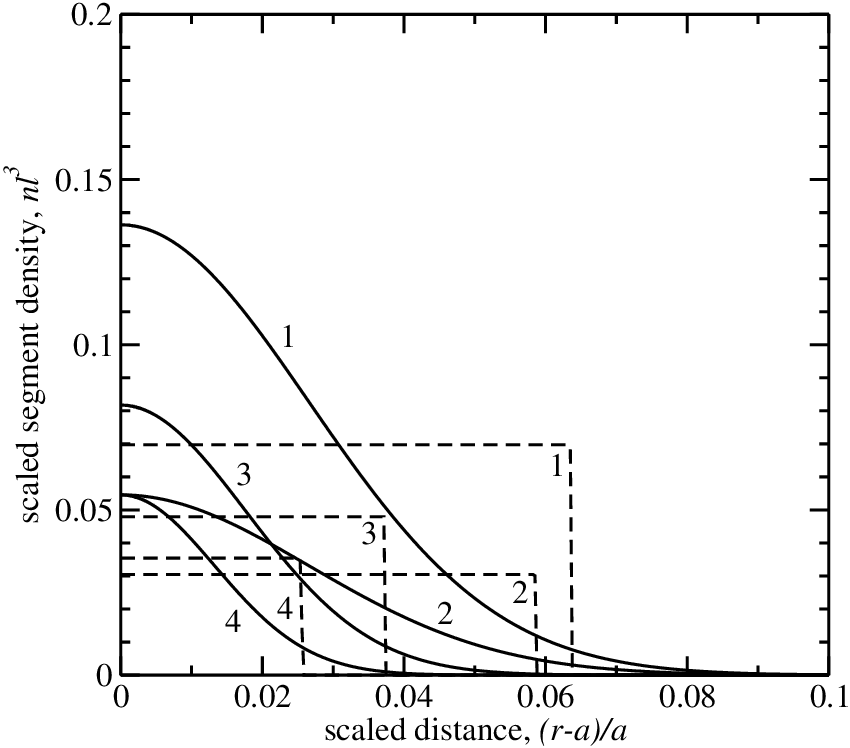}
  \caption{\label{fig:thinneutraldensity} The (scaled) segment density
    distributions $\phi(r) = n(r) l^3$ as a function of the (scaled)
    distance from the bare surface $(r - a) / a$ for the particles
    whose electrophoretic mobilities are shown in
    figure~\ref{fig:thinneutralmobility}. The density and thickness of
    the uniform layers ({\em dashed} lines) have been adjusted 
    to have ($i$) the same hydrodynamic layer thickness $L_h$ and
    ($ii$) the same number of polymer segments $N \sigma_p$ as the
    corresponding non-uniform layer. See
    table~\ref{tab:thinneutraltable} for the coating parameters, and
    figure~\ref{fig:thinneutralmobility} for the electrophoretic
    mobilities.}
\end{figure}

Let us consider how either $n_1$ and $\delta$ or $n_0$ and $L$ are
related to the electrophoretic mobility and hydrodynamic layer
thickness $L_h$, both of which are routinely measured. Because $L_h$
depends on the permeability $\bsl^2 = 1/(6 \pi n(r) a_s)$, the Stokes
radius of the segments $a_s$ must be correctly specified; here, $a_s =
0.175$~\AA \ is a value for poly(ethylene glycol) (PEG) that provides
a good `fit' of theory (with a self-consistent mean-field description
of the polymer density distribution) to measured electrophoretic
mobilities~\citep{Hill:2004a}. The physical characteristics of four
representative coatings, each decorating the surface of an impermeable
spherical colloid with surface charge density $\sigma_c \approx
-0.78~\mu$Ccm$^{-2}$ and radius $a = 500$~nm are listed in
table~\ref{tab:thinneutraltable}. Each of the four sets of parameters
(cases 1--4), which specify $L_h$ and $N$ (columns 2 and 3), have been
used to characterize Gaussian-like {\em and} uniform (step-like)
segment density distributions. Note that the hydrodynamic layer
thicknesses were established iteratively from solutions of problem (U)
at high ionic strength. Clearly, the various molecular weights and
hydrodynamic coating thicknesses lead to a variety of physical
thicknesses ($\delta$ and $L$) and densities ($n_0$ and $n_1$). As
shown in figure~\ref{fig:thinneutraldensity}, the extent (maximum
density) of each non-uniform layer is significantly greater (less)
than that of its uniform counterpart.

  \begin{table}
    \begin{center}
      \caption{\label{tab:thinneutraltable} Parameters that
	characterize the layers whose segment density distributions
	and electrophoretic mobilities are shown in
	figures~\ref{fig:thinneutraldensity}
	and~\ref{fig:thinneutralmobility}: $\sigma_p l^2 = 0.072$ with
	$l=0.71~$nm; $a_s = 0.175$~\AA; $a =500$~nm; $\sigma_c =
	-0.780~\mu$Ccm$^{-2}$.}
      \begin{tabular*}{\columnwidth}{@{\extracolsep{\fill}}cccccccc}
	\hline
\multicolumn{3}{c}{} & \multicolumn{2}{c}{Gaussian-like}& \multicolumn{3}{c}{Step-like}\\ \cline{4-5} \cline{6-8}
	case & $L_h$~(nm) & $N$ & $n_1$~(M) & $\delta$~(nm)& $n_0$~(M) & $L$~(nm) & $\bsl$ (nm)\\ \hline
	1 & 28 & 46.2 & 0.63 & 18.8 & 0.32 & 31.8 & 3.9 \\
	2 & 23 & 18.5 & 0.25 & 18.8 & 0.14 & 29.3 & 6.0 \\
	3 & 14 & 18.2 & 0.38 & 12.5 & 0.22 & 18.7 & 4.8 \\ 
	4 & 7.4& 9.10 & 0.25 & 9.4  & 0.16 & 12.8 & 5.6 \\ \hline
      \end{tabular*}
    \end{center}
  \end{table}   

\begin{figure}
\centering
\subfigure[]{\includegraphics[height=6cm]{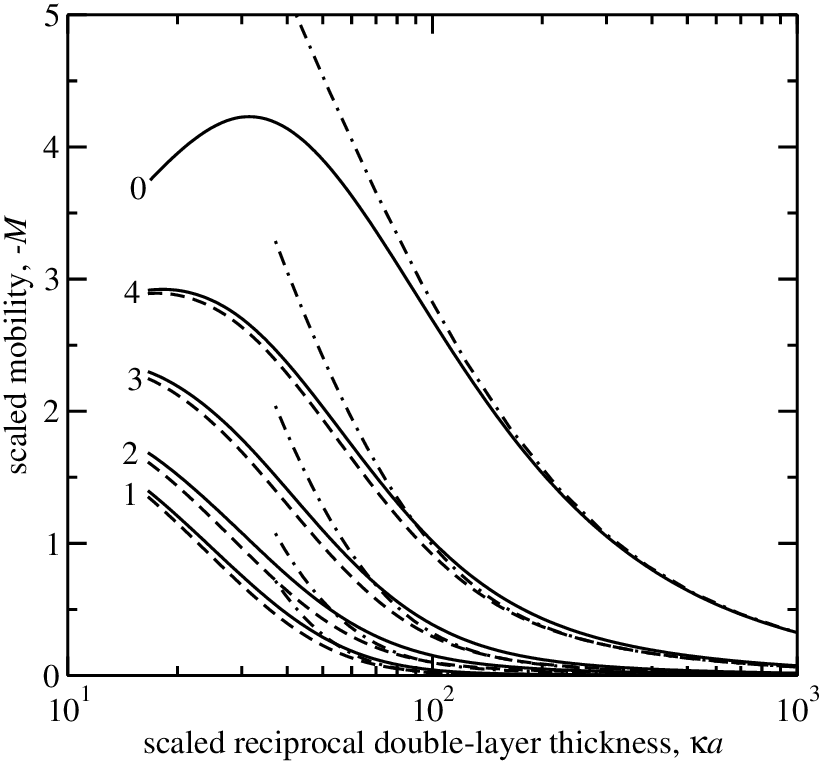}}
\subfigure[]{\includegraphics[height=6cm]{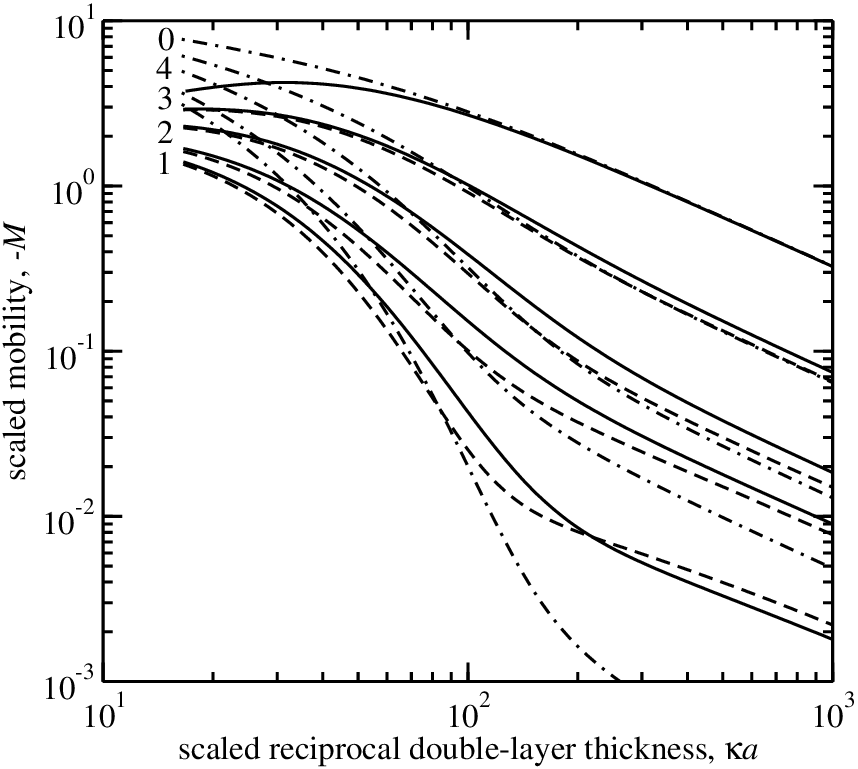}}
  \caption{\label{fig:thinneutralmobility} The (scaled)
    electrophoretic mobility $M = 3 \eta e V / (2 \epsilon_s
    \epsilon_o \kb T E)$ as a function of the (scaled) reciprocal
    double-layer thickness $\ka$ (aqueous NaCl at $T = 25$\degc \ with
    radius $a = 500$~nm). The surface charge density $\sigma_c \approx
    -0.780~\mu$Ccm$^{-2}$ at all ionic strengths. {\em Solid} lines
    are the full model with non-uniform segment density distributions
    ({\em solid} lines in figure~\ref{fig:thinneutraldensity}), {\em
    dashed} lines are the full model with uniform (step-like) segment
    density distributions ({\em dashed} lines in
    figure~\ref{fig:thinneutraldensity}), and {\em dash-dotted} lines
    are Ohshima's theory~\citep[][Eqn.~(11.4.27)]{Ohshima:1995}. See
    table~\ref{tab:thinneutraltable} for the coating parameters (cases
    1--4, with case 0 indicating the bare particle), and
    figure~\ref{fig:thinneutraldensity} for the segment density
    distributions.}
\end{figure}

The electrophoretic mobilities are shown in
figure~\ref{fig:thinneutralmobility} over a range of ionic
strengths. Despite large variations in density and extent,
constraining each uniform layer ({\em solid} lines) to have the same
hydrodynamic thickness as its respective non-uniform layer ({\em
dashed} lines) leads to similar mobilities. This is to be expected at
low ionic strength, but the close correspondence at high ionic
strength is surprising.  When the diffuse double layer resides mostly
within the layers (at high ionic strength), the mobility is expected
to be more sensitive to details of the distribution. However, when the
densities and thicknesses are constrained by
Eqn.~(\ref{eqn:densityconstraint}), the mobility reflects---to a good
approximation---the hydrodynamic layer thickness. Clearly, if the
hydrodynamic layer thickness is known, the electrophoretic mobility
has a remarkably close connection to the underlying surface charge
over a wide range of ionic strengths.

The logarithmically scaled mobility axis (panel (b) in
figure~\ref{fig:thinneutralmobility}) facilitates a closer comparison
of the numerically exact solutions with Ohshima's well-known
analytical approximation for thin uniform
layers~\citep[][Eqn.~(11.4.27)]{Ohshima:1995} ({\em dash-dotted}
lines). Good agreement is seen when $L / a < 0.02$ and $\ka > 100$,
but, as expected, the approximate theory breaks down at low ionic
strength. As shown below, Ohshima's more complicated
formula~\citep[][Eqn.~(11.4.24)]{Ohshima:1995} can be advantageous
with higher electrostatic potentials. Here, however, the surface
charge density is low enough for the breakdown to be attributed to the
finite particle size, \ie, to polarization and relaxation.

\subsection{Thin coatings and passage to the high-ionic-strength limit} \label{sec:highionicstrength}

\begin{figure}
\centering
\subfigure[]{\includegraphics[height=6cm]{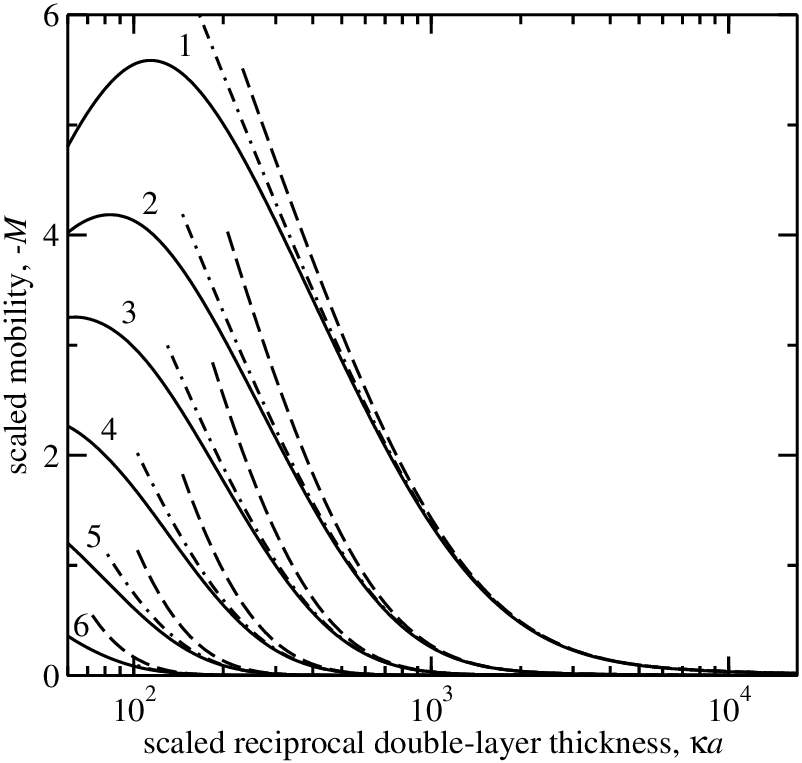}}
\subfigure[]{\includegraphics[height=6cm]{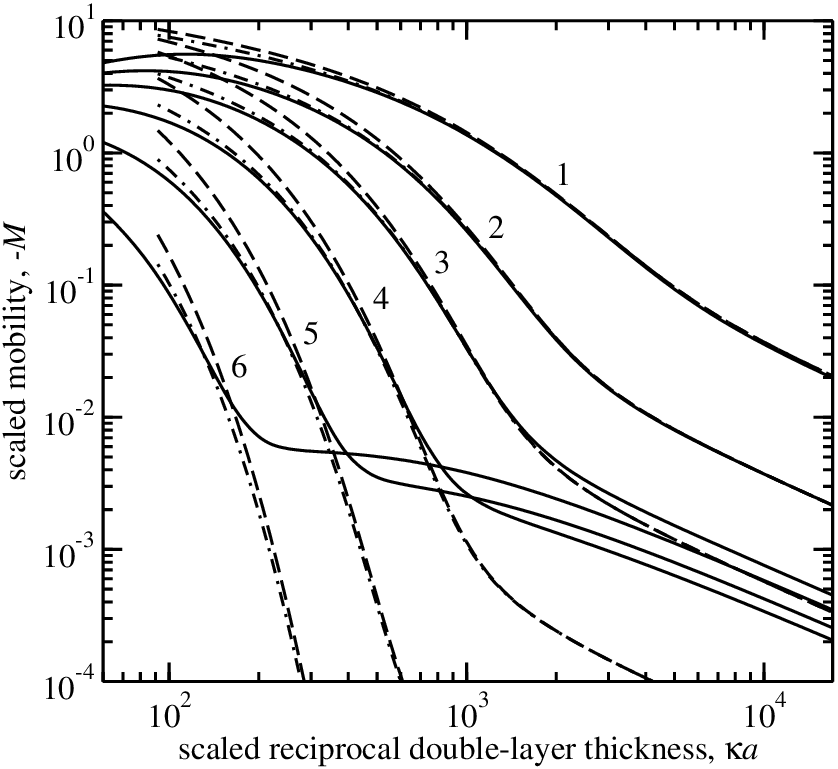}}
\caption{\label{fig:smallkalarge} The (scaled) electrophoretic
  mobility $M = 3 \eta e V / (2 \epsilon_s \epsilon_o \kb T E)$ as a
  function of the (scaled) reciprocal double-layer thickness $\ka$
  (aqueous NaCl at $T = 25$\degc \ with radius $a = 1750$~nm). The
  surface charge density $\sigma_c \approx -1.95~\mu$Ccm$^{-2}$ at all
  ionic strengths. {\em Solid} lines are the full model with uniform
  segment density distributions, and the {\em dashed} and {\em
  dash-dotted} lines are Ohshima's theory for
  low~\citep[][Eqn.~(11.4.27)]{Ohshima:1995} and
  arbitrary~\citep[][Eqn.~(11.4.24)]{Ohshima:1995} electrostatic
  potentials. See table~\ref{tab:smallkalarge} for the coating
  parameters.}
\end{figure}

Surprisingly, the full model deviates from Ohshima's flat-plate
theories at high ionic strength, even with very thin layers. For
example, figure~\ref{fig:smallkalarge} shows the electrophoretic
mobilities of particles whose (uniform) coatings are all very thin
compared to the particle radius ($a = 1750$~nm). The coating
parameters and various dimensionless ratios are listed in
table~\ref{tab:smallkalarge}. Note that all the layers (cases 1--6)
have the same grafting density and polymer molecular weight, so the
increase in thickness $L$ is accompanied by a decrease in density
$n_0$ and, hence, an increase in permeability $\bsl^2$. Nevertheless,
Brinkman screening ensures that the hydrodynamic layer thickness $L_h$
increases with $L$.

  \begin{table}
    \begin{center}
      \caption{\label{tab:smallkalarge} Parameters that characterize
	the uniform layers on particles whose electrophoretic
	mobilities are shown in figure~\ref{fig:smallkalarge}:
	$\sigma_p l^2 = 0.072$ with $l=0.71~$nm; $N = 100$; $a_s =
	0.175$~\AA; $a = 1750$~nm; $\sigma_c \approx
	-1.95~\mu$Ccm$^{-2}$.}
      \begin{tabular*}{\columnwidth}{@{\extracolsep{\fill}}cccccccc}
	\hline case & $L_h$~(nm) & $n_0$~(M) & $L$~(nm) & $\bsl$ (nm) & $\bsl/L$ & $L/a$\\ \hline 
	1 & 1.10 & 14.0 & 1.69 & 0.591 & 0.35 & $6.3 \times 10^{-4}$\\
	2 & 4.37 & 4.34 & 5.45 & 1.07 &  0.20 & $2.7 \times 10^{-3}$\\
	3 & 8.75 & 2.31 & 10.2 & 1.47 & 0.14 & 0.005\\
	4 & 17.5 & 1.20 & 19.6 & 2.04 & 0.10 & 0.01\\
	5 & 35.0 & 0.613 & 37.9 & 2.86 & 0.075 & 0.02\\
	6 & 70.0 & 0.307 & 74.0 & 4.04 & 0.055 &0.04\\ \hline
      \end{tabular*}
    \end{center}
  \end{table}   

Despite a substantial decrease in $L / a$---relative to the cases
presented in figure~\ref{fig:thinneutralmobility}---there is a
significant difference between the approximate and `exact' results
when $L / a > 0.01$. Remarkably, at high ionic strength, the mobility
decreases slowly with decreasing layer thickness, whereas the
flat-plate theories ({\em dashed} and {\em dash-dotted} lines) infer a
monotonic decline, yielding much lower mobilities. We attribute this
unexpected behavior to the finite particle size and a transition to
the regime where $\kappa \bsl \gg 1$ with $\bsl / L \ll 1$. Note that
the pressure gradient must be identically zero with a perfectly flat
interface. However, when $\kappa L \gg 1$, fluid in the electrically
neutral region of the polymer ($\kappa^{-1} <r - a < L$) must be
driven by viscous stresses when $\kappa \bsl > 1$. With a finite
radius of curvature, radial and tangential pressure gradients then
develop to move fluid through the polymer, into and out of the
underlying diffuse double layer. As demonstrated by Hill, Saville and
Russel~\citep[][Fig.~6]{Hill:2003a}, decreasing the permeability
increases the tendency of fluid to enter and exit the polymer layer
radially, and this, presumably, decreases the radial and tangential
pressure fluctuations.

As shown in figure~\ref{fig:smallkalarge}, the breakdown of the
flat-plate approximation at high ionic strength can be significant for
neutral layers with moderate thickness and permeability. For example,
the mobility with $L \approx 20$~nm and $\bsl \approx 2$~nm (case 5)
deviates from Ohshima's theory when the ionic strength exceeds 18~mM
($\ka > 770$). Note that there exists an intermediate range of ionic
strengths, with $\kappa \bsl$ sufficiently small and $\ka$
sufficiently large, where Ohshima's theory provides an excellent
approximation. As expected, the higher surface charge here---relative
to the cases presented in
figure~\ref{fig:thinneutralmobility}---demonstrates that Ohshima's
formula~\citep[][Eqn.~(11.4.24)]{Ohshima:1995} ({\em dash-dotted}
lines) for arbitrary potentials can indeed provide a significantly
better approximation than his simpler
formula~\citep[][Eqn.~(11.4.27)]{Ohshima:1995} for low potentials
({\em dashed} lines).

\subsection{Polarization and relaxation, and the influence of particle size} \label{sec:polarization}

\begin{figure}
\centering \includegraphics[height=5cm]{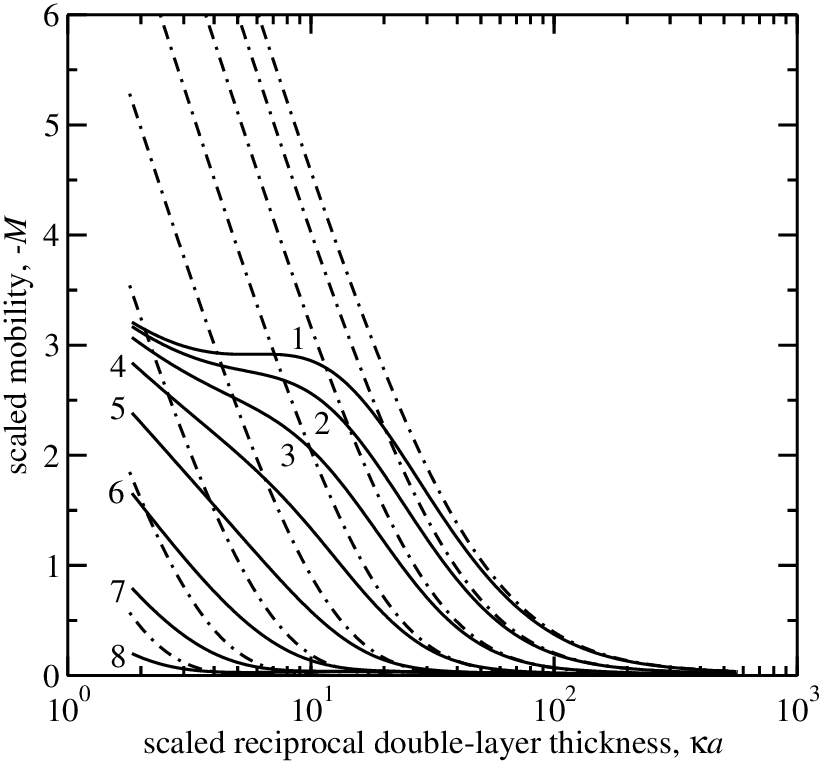}
\hspace{1cm} \includegraphics[height=5cm]{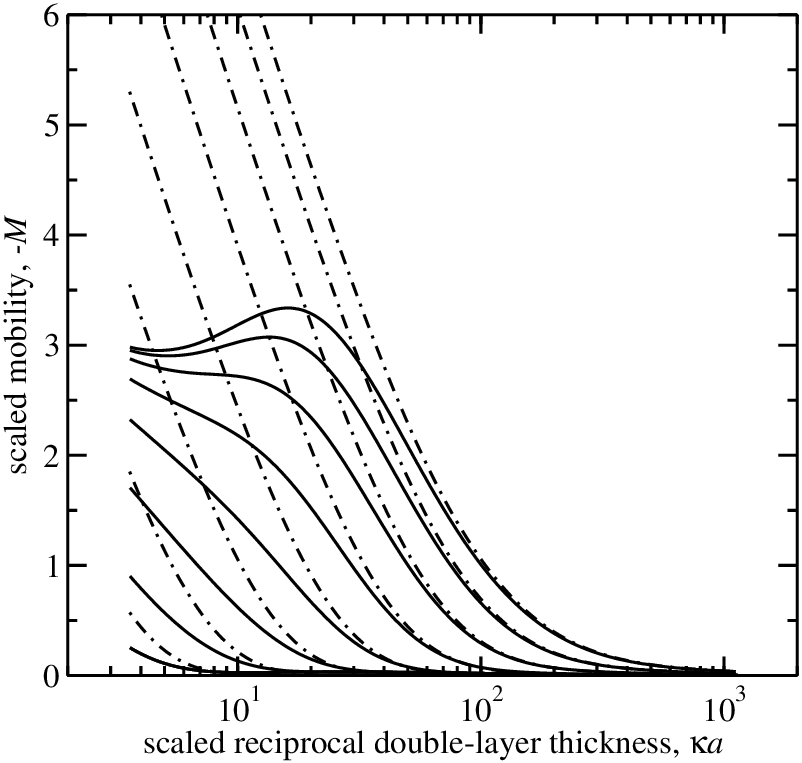}\\
 \vspace{1cm} \includegraphics[height=5cm]{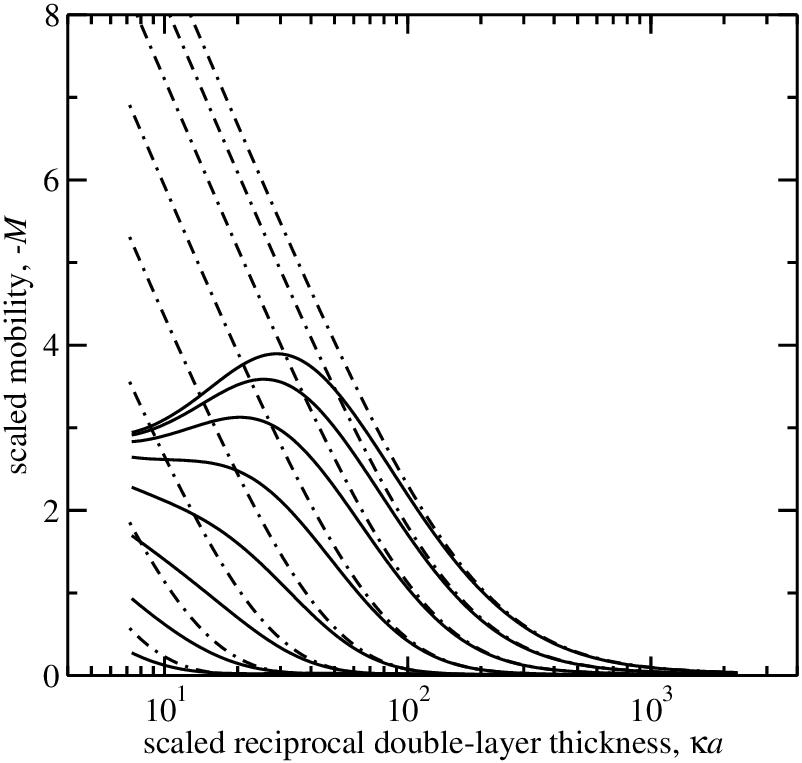} \hspace{1cm}
\includegraphics[height=5cm]{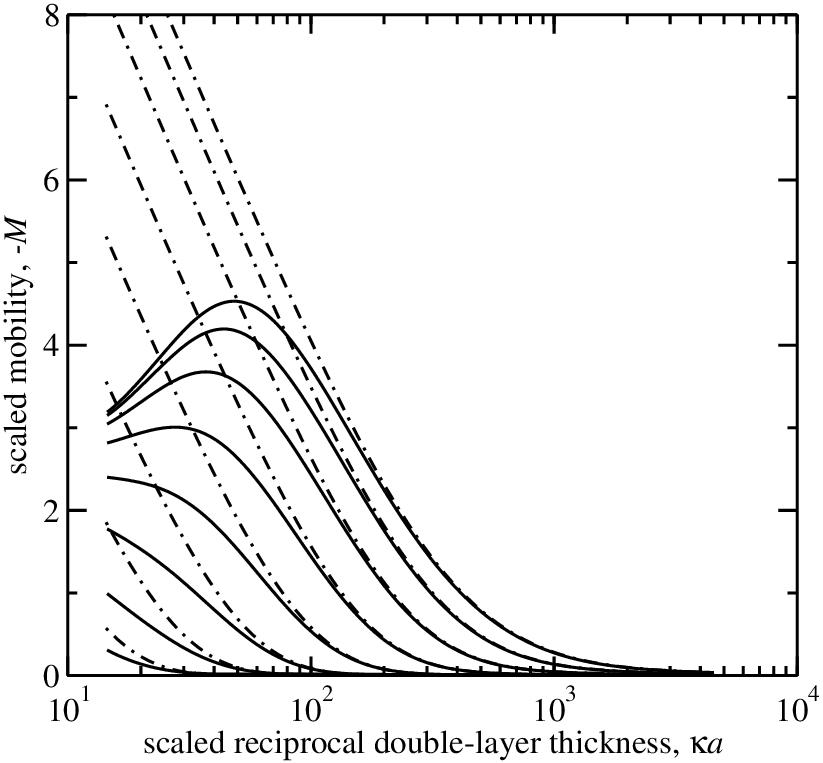}\\
 \vspace{1cm} \includegraphics[height=5cm]{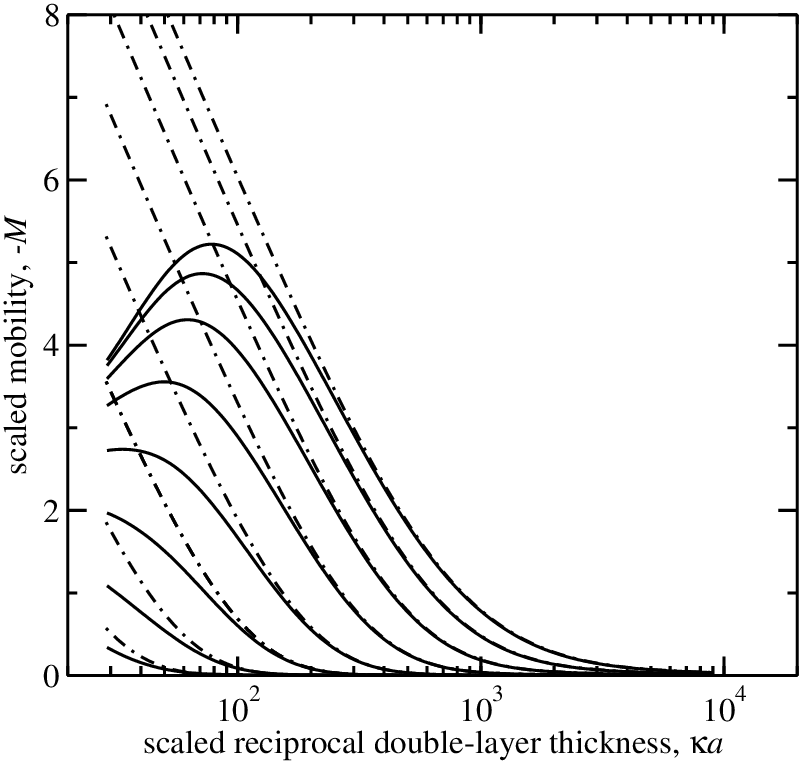} \hspace{1cm}
\includegraphics[height=5cm]{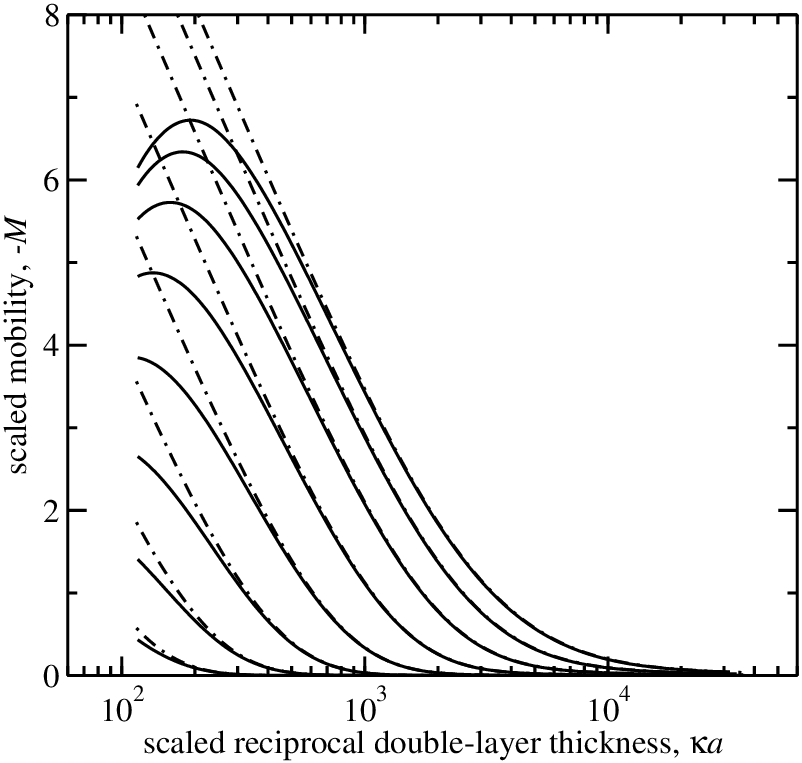}
  \caption{\label{fig:ohshimacompb} The (scaled) electrophoretic
    mobility $M = 3 \eta e V / (2 \epsilon_s \epsilon_o \kb T E)$ as a
    function of the (scaled) reciprocal double-layer thickness $\ka$
    (aqueous NaCl ($I = 10^{-4}$--$10$~M) at $T = 25$\degc \ with
    radius $a \approx 55, 109, 219, 438, 875$ and 3500~nm ({\em
    top-left} to {\em bottom-right})): $\sigma_p l^2 = 0.072$ with
    $l=0.71~$nm; $N = 100$; $a_s = 0.175$~\AA; $\sigma_c \approx
    -1.95~\mu$Ccm$^{-2}$. The particles have uniform neutral coatings
    with thicknesses $L$ and densities $n_0$ specified according to
    Eqn.~(\ref{eqn:constraint}) with $F_s = 1$ and $L_h \approx (17.5
    / 32) 2^\nu$~nm, where $\nu=1$, 2, 3,..., 8 (labeled in the first
    panel). {\em Solid} lines are the full model with uniform segment
    density distributions, and the {\em dash-dotted} lines are
    Ohshima's theory~\citep[][Eqn.~(11.4.24)]{Ohshima:1995}.}
\end{figure}

\begin{figure}
\centering
\includegraphics[height=5cm]{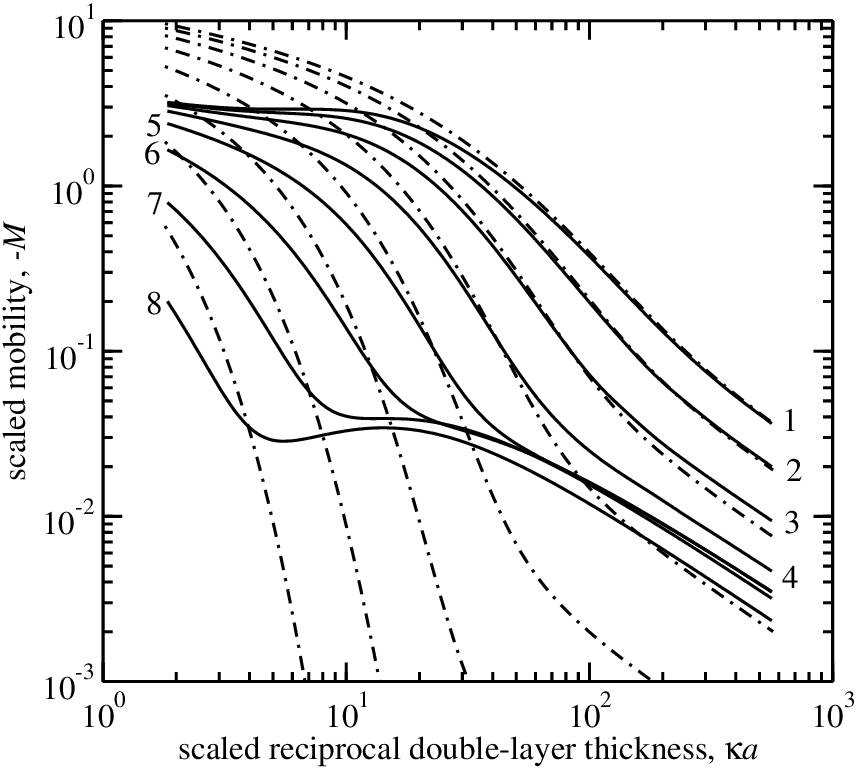} \hspace{1cm}
\includegraphics[height=5cm]{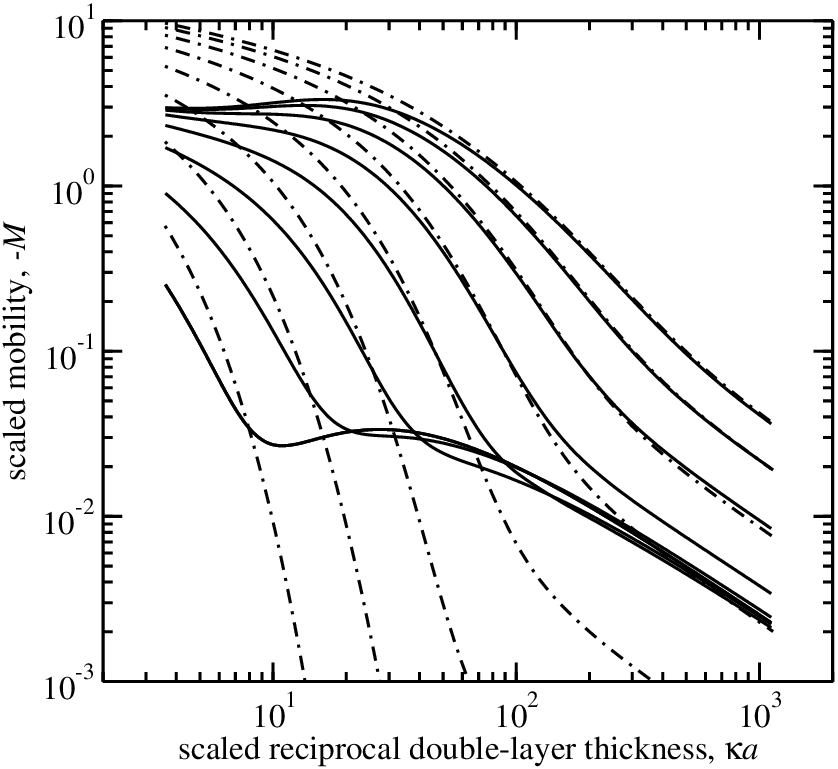}\\
\vspace{1cm}    \includegraphics[height=5cm]{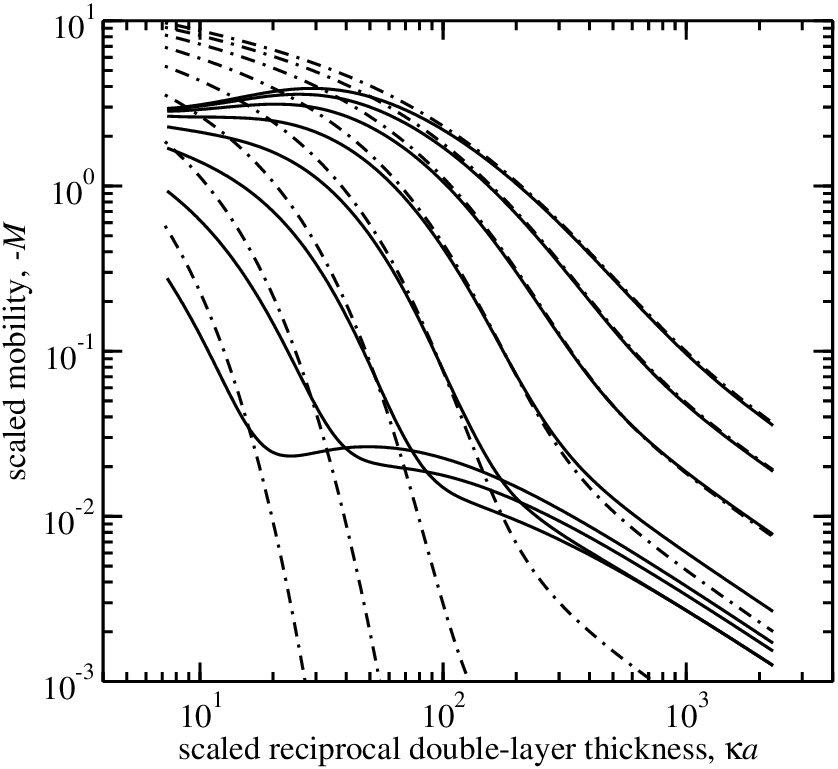} \hspace{1cm}
\includegraphics[height=5cm]{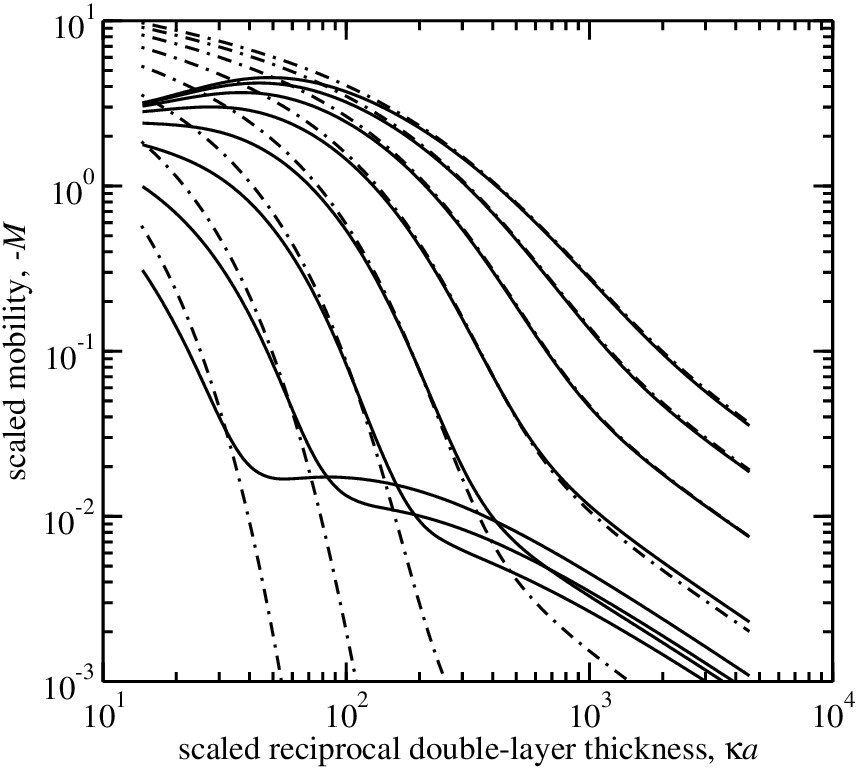}\\
\vspace{1cm}    \includegraphics[height=5cm]{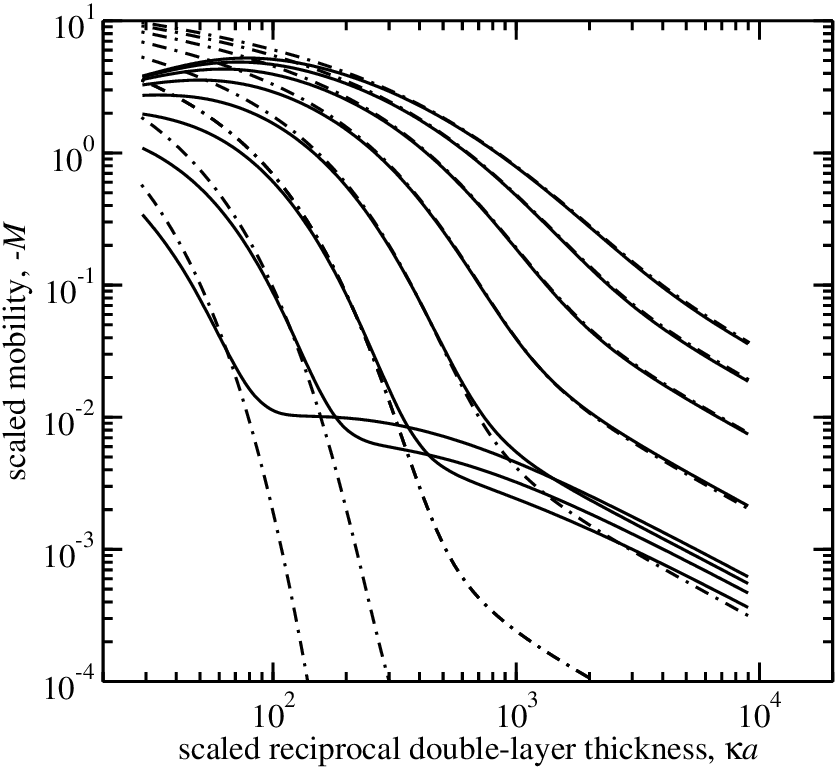} \hspace{1cm}
\includegraphics[height=5cm]{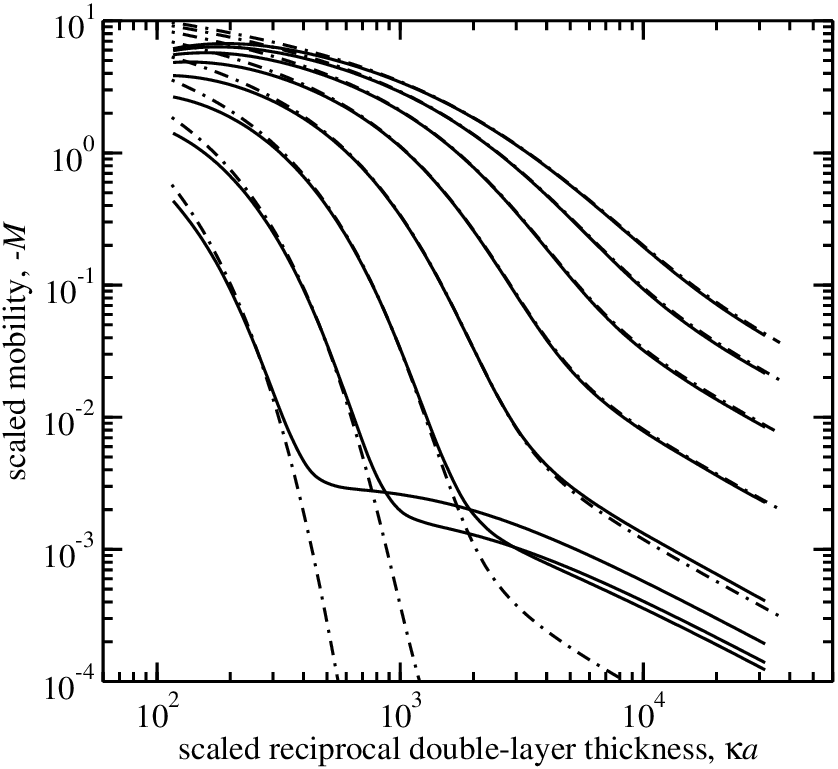}
\caption{\label{fig:ohshimacomp} As in
  figure~\ref{fig:ohshimacompb}, but with logarithmically scaled
  mobility axes.}
\end{figure}

With polymer layers, clearly identifying the regions of the parameter
space where polarization and relaxation are significant is not
straightforward.  The calculations presented in
figures~\ref{fig:ohshimacompb} and~\ref{fig:ohshimacomp}
systematically explore the influence of particle size~($a \approx
50$--3500~nm) and layer thickness. Similarly to
figure~\ref{fig:smallkalarge} with $a = 1750$~nm, all the coatings
have the same grafting density ($\sigma_p l^2 = 0.072$) and polymer
molecular weight ($N =100$), so the permeabilities vary with the
(specified) {\em hydrodynamic} thickness {\em and} particle
radius. The numerically exact results ({\em solid} lines), which are
compared to Ohshima's theory for arbitrary surface
potentials~\citep[][Eqn.~(11.4.24)]{Ohshima:1995} ({\em dash-dotted}
lines), demonstrate how particle size influences the mobility. Because
the surface charge densities are the same, varying the surface
curvature is accompanied by a change in the surface potential at fixed
ionic strength ($\kappa \sim \sqrt{I}$). Increasing the coating
thickness decreases the mobility, mimicking the effect of decreasing
the charge. As expected for bare particles, decreasing the radius
increases polarization, which, in turn, decreases the mobility and
shifts the maxima to smaller values of $\ka$.

Neutral coatings attenuate polarization, however, as demonstrated by
the increasing accuracy of Ohshima's theory with increasing layer
thickness when $\ka \approx 300$ and $a = 3500$~nm (last panel in
figure~\ref{fig:ohshimacomp}).  The close correspondence for the
thickest coating ($L / a \approx 0.02$) establishes the validity of
the flat-plate approximation and the negligible influence of
polarization under these conditions. It follows that decreasing the
layer thickness should improve the flat-plate approximation. However,
the agreement clearly diminishes, and this must be attributed to
polarization and relaxation. Note that decreasing the layer thickness
(with $\ka = 300$ and $a = 3500$~nm) yields $\kappa L = 1$ when $L
\approx 12$~nm. Therefore, the four thinnest layers have $\kappa L <
1$ with $\ka \gg 1$, this being the regime in which
Saville~\citep{Saville:2000} developed a semi-analytical theory to
highlight the role of polarization and relaxation. Comparing the exact
results with Ohshima's theory (last panel in
figure~\ref{fig:ohshimacomp}) confirms that polarization and
relaxation can be significant when $\ka$ is large. Clearly, when most
of the double layer resides beyond the polymer, \ie, when $\kappa L
\ll 1$, the ratio of polarization (by convection and electromigration)
to relaxation (by molecular diffusion) is expected to be maximal.

\subsection{Interpreting experiments} \label{sec:interpretation}

As shown in section~\ref{sec:thinneutralcoatings}, the full
electrokinetic model provides satisfactory predictions of the
electrophoretic mobility with uniform layers when the coating density
and thickness reflect the correct adsorbed amount $\sigma_p N$
(according to Eqn.~(\ref{eqn:uniform})) {\em and} the `correct'
hydrodynamic thickness $L_h$. This approach requires the Stokes radius
of the polymer segments $a_s$ to be specified. Note that, in the
Brinkman model, the designation of monomer or statistical units to a
polymer segment is arbitrary, as long as ($i$) the number of segments
is unambiguously related to the molecular weight, \eg, $W = N M_s =
N_m M_m$ with $l_c = N l = N_m l_m$, and ($ii$) the Stokes radius of
the segments is appropriate.

In general, determining the hydrodynamic thickness of a layer requires
a numerical solution of Brinkman's equations or, equivalently, a
solution of the (U) problem at high ionic strength. The results in
table~\ref{tab:thinneutraltable} for uniform layers demonstrate that
the hydrodynamic thickness
\begin{equation}
  L_h \approx L - \bsl,
\end{equation}
where $\bsl^{-2} = 6 \pi a_s F_s n_0$. It follows from
Eqn.~(\ref{eqn:uniform}) that
\begin{equation} \label{eqn:constraint}
  (L_h - L)^2 \approx \frac{(1+L/a)^3-1}{18 \pi a_s F_s N \sigma_p / a},
\end{equation}
which is a convenient and useful
formula\footnote{Eqn.~(\ref{eqn:constraint}) is easily solved
iteratively, beginning with $L_h = L$.} relating the thickness
$L$ of a uniform layer to the physical parameters $L_h$, $N \sigma_p$,
$a_s$ and $a$. The approximation is accurate when the $\bigO{\eta M E
/ \bsl}$ viscous force (per unit volume) on the fluid in the layer is
small relative to the $\bigO{ \eta M E L / \bsl^2}$ frictional (Darcy)
drag force, \ie, when $\bsl / L < 1$.

Equation~(\ref{eqn:constraint}) can be used to reduce the number of
unknown degrees of freedom when interpreting electrokinetic
experiments. Consider, for example, Cohen and Khorosheva's experiments
with poly(ethylene glycol) (PEG) terminally anchored to the surface of
multilamellar liposomes with $a \approx
1750$~nm~\citep{Cohen:2001}. Because the PEG grafting density, surface
charge density, and PEG molecular weight are `known', Cohen and
Khorosheva devised a very simple electrokinetic model to infer a value
of $L$ and $\bsl$ for each of the five molecular weights of PEG, \ie,
ten `fitting' parameters. However, adopting
Eqn.~(\ref{eqn:constraint}) reduces the number of such parameters to
six: one value of $L$ for each molecular weight, and one value of
$a_s$.

Recently, Hill interpreted Cohen and Khorosheva's experiments using a
self-consistent mean-field model of the polymer segment density
distribution~\citep{Hill:2004a}. This reduced the problem to one of
determining $a_s \approx 0.175$~\AA \ alone. The remarkably small
Stokes radius established for PEG statistical segments with (Kuhn)
length $l = 0.71$~nm mimics the (negative) correlations that exist
between micro-scale segment density and fluid velocity fluctuations,
which are not explicitly accounted for in the Brinkman
model~\citep{Hill:2004a}. Physically, regions of low segment density
favor high fluid velocities, thereby reducing the average drag force
on an otherwise statistically homogeneous porous medium. This suggests
that the Stokes radius may be a complicated function of the polymer
structure and density. In relatively dense cross-linked gels, for
example, density-velocity correlations may be much weaker than in
brush-like layers of relatively loosely confined polymer. This would
lead to a Stokes radius that is much more representative of the
segment physical size.

\begin{figure}
\centering
\vspace{1cm}
\includegraphics[height=7cm]{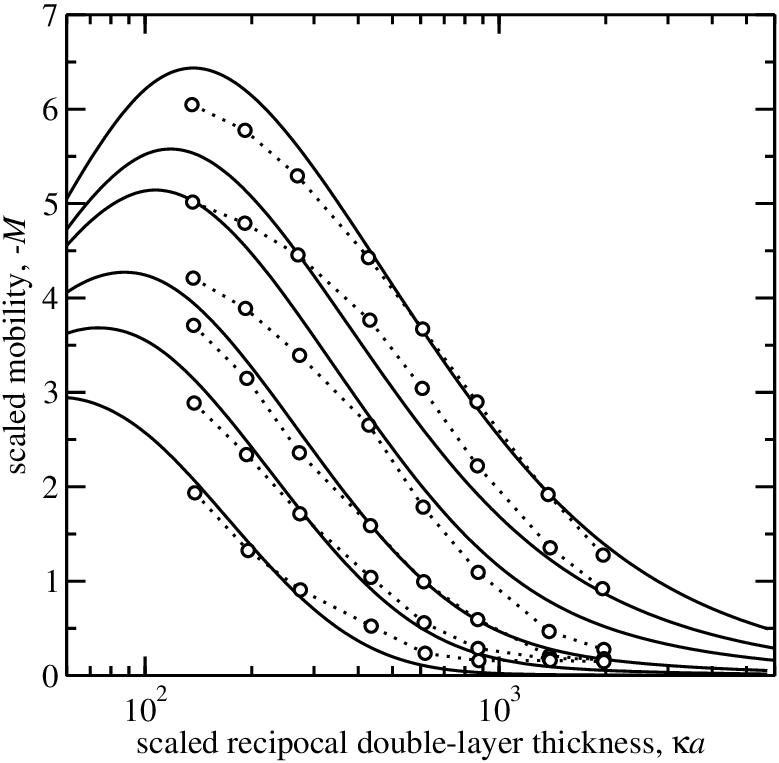}
  \caption{\label{fig:1750nmliposomesmobility} The (scaled)
    electrophoretic mobility $M = 3 \eta e V / (2 \epsilon_s
    \epsilon_o \kb T E)$ of spherical liposomes with coatings of
    terminally anchored PEG as a function of the (scaled) reciprocal
    double-layer thickness $\ka$ (aqueous NaCl at $T = 25$\degc \ with
    radius $a = 1750$~nm) for various numbers of statistical segments
    per PEG chain $N = 0$, 4.5, 14, 28, 42, and 70 (increasing
    downward); the molecular weights of the PEG chains are,
    respectively, $W = 0$, 0.35, 1, 2, 3 and 5~kgmol$^{-1}$. The
    surface charge density $\sigma_c \approx -1.95~\mu$Ccm$^{-2}$ at
    all ionic strengths. {\em Circles} (with dotted lines to guide the
    eye) are Cohen and Khorosheva's experimental
    data~\citep{Cohen:2001}, and {\em solid} lines are the full model
    with uniform segment density distributions, whose density and
    thickness are chosen to yield the same hydrodynamic layer
    thicknesses as Hill's self-consistent mean-field description of
    the layers~\citep{Hill:2004a}. See
    table~\ref{tab:1750nmliposomestable} for parameters.}
\end{figure}

To demonstrate the application of Eqn.~(\ref{eqn:constraint}), we
adopt hydrodynamic thicknesses $L_h$ and the segment Stokes radius
$a_s \approx 0.175$~\AA \ as established by Hill's self-consistent
mean-field description of the polymer~\citep{Hill:2004a}, and compare
the resulting mobilities with Cohen and Khorosheva's experimental
data. As shown in figure~\ref{fig:1750nmliposomesmobility}, the
electrophoretic mobilities predicted in this manner are in good
agreement with the measured
values. Table~\ref{tab:1750nmliposomestable} summarizes the parameters
for each polymer molecular weight. Note that all the coatings have the
same grafting density. Comparing the actual hydrodynamic thicknesses
of the uniform layers (column 6) with the desired values (column 2)
demonstrates that the approximation underlying
Eqn.~(\ref{eqn:constraint}) is accurate when $\bsl / L \ll 1$, \ie,
when the PEG molecular weight is greater than 1~kgmol$^{-1}$ at this
grafting density. The results are also consistent with expectations
that the density and, hence, permeability of polymer brushes are
independent of the molecular weight at a fixed grafting
density. Clearly, the Brinkman screening length decreases with
increasing molecular weight, approaching $\bsl \approx 2$~nm. The
layer thickness increases linearly with molecular weight, yielding $L
\approx (0.21 + 0.22 N) l$ ($l \approx 0.71$~nm) when $W >
1$~kgmol$^{-1}$. As expected, this formula is the same as Hill's for
the high-molecular-weight limit of the hydrodynamic layer thickness,
$L_h \sim 0.22 N l$, at this grafting density.

  \begin{table}
    \begin{center}
      \caption{\label{tab:1750nmliposomestable} Parameters
	characterizing the polymer layers in Cohen and Khorosheva's
	(\CK) experiments with terminally anchored PEG on
	multilamellar liposomes~\citep{Cohen:2001}. The hydrodynamic
	thickness of the coatings $L_h$ and the Stokes radius of the
	segments $a_s \approx 0.175$~\AA \ are specified from Hill's
	calculations with a self-consistent mean-field description of
	the polymer~\citep{Hill:2004a}. These parameters characterize
	uniform layers with thickness $L$ and permeability $\bsl^2$,
	specified according to Eqn.~(\ref{eqn:constraint}).}
      \begin{tabular*}{\columnwidth}{@{\extracolsep{\fill}}ccccccc}
	\hline $W$~(gmol$^{-1}$) & $N$ & $L_h$~$^1$~(nm) & $L$~(nm)
	& $n_0$~(M)&$\bsl$~(nm) & $L_h$~$^2$~(nm)\\ \hline
	350 & 4.5 & 0.42 & 5.5 & 0.19 & 5.1 & 1.5\\
	1000 & 14 & 2.0 & 4.7 & 0.71 & 2.7 & 2.2\\
	2000 & 28 & 4.3 & 6.5 & 1.0 & 2.2 & 4.3\\
	3000 & 42 & 6.5 & 8.6 & 1.2 & 2.1 & 6.5\\
	5000 & 70 & 11 & 13 & 1.3 & 2.0 & 11\\ \hline
      \end{tabular*}
    \end{center}
	{\small $^1$~Actual or desired values from Hill's
	  self-consistent mean-field calculations~\citep{Hill:2004a}.\\
      $^2$~Actual values for uniform layers.}

  \end{table}   

\subsection{Terminally anchored PEG on lipid bilayer membranes (stealth liposomes)} \label{sec:peg}

Cohen and Khorosheva's experiments, introduced above, were performed
with relatively large multilamellar liposomes. This permitted the
particles to be observed under magnification in their electrophoresis
device. However, stealth liposomes used for drug delivery have much
smaller radii $ a \sim 100$~nm. With the same surface charge, polymer,
and polymer grafting density, the smaller radius is expected to change
the surface potential and, possibly, vary the polymer segment density
distribution. Therefore, with knowledge of the Stokes radius for PEG
(Kuhn) segments, obtained from Cohen and Khorosheva's experiments with
$a = 1750$~nm, we can predict the electrophoretic mobilities of {\em
therapeutic} stealth liposomes with $a = 100$~nm, for example.

\begin{figure}
  \centering
  \vspace{1cm}
  \includegraphics[height=7cm]{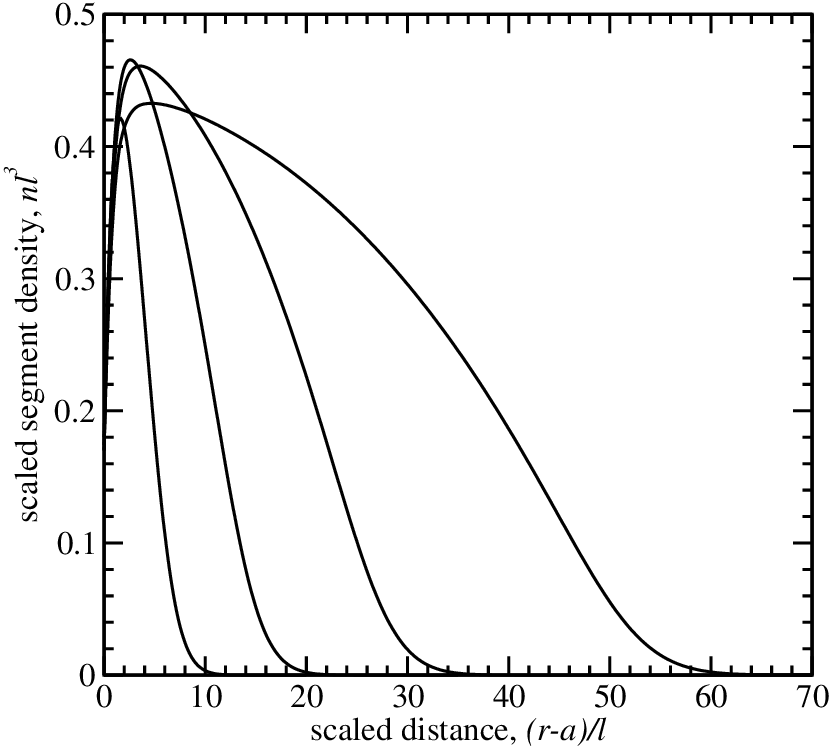}
  \caption{\label{fig:100nmliposomesdensity} The (scaled) radial
    segment density distribution $\phi = n(r) l^3$ of PEG chains
    terminally anchored to the surface of spherical (therapeutic)
    liposomes with radius $a = 100$~nm as a function of the (scaled)
    distance from the surface of the bare liposome $(r - a)/l$. The
    data are for representative numbers of statistical segments per
    chain $N = 28$, 70, 140 and 280, with molecular weights $W \approx
    2$, 5, 10 and 20~kgmol$^{-1}$. The self-consistent mean-field
    potential is specified according to Hill~\citep{Hill:2004a} with
    $v_2 / v_1 = 3.3$, $v_2 / l^3 = 0.27$, $\sigma_p l^2 = 0.072$ and
    $l = 0.71$~nm.}
\end{figure}

\begin{figure}
\centering
\vspace{1cm}
  \includegraphics[height=7cm]{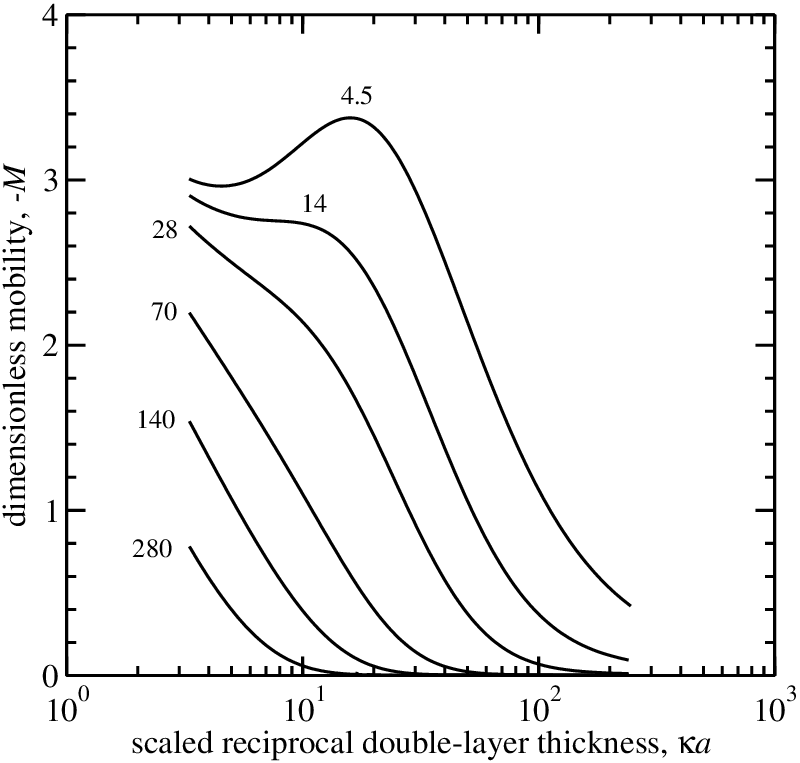}
  \caption{\label{fig:100nmliposomesmobility} The (scaled)
    electrophoretic mobility $M = 3 \eta e V / (2 \epsilon_s
    \epsilon_o \kb T E)$ of spherical liposomes with coatings of
    terminally anchored PEG as a function of the (scaled) reciprocal
    double-layer thickness $\ka$ (aqueous NaCl at $T = 25$\degc \ with
    radius $a = 100$~nm) for various numbers of statistical segments
    per PEG chain $N = 4.5$, 14, 28, 70, 140 and 280; the molecular
    weights of the PEG chains are, respectively, $W \approx 0.35$, 1,
    2, 5, 10 and 20~kgmol$^{-1}$. The surface charge density $\sigma_c
    \approx -1.95~\mu$Ccm$^{-2}$ at all ionic strengths. These
    predictions are from the full model with a self-consistent
    mean-field description of the PEG segment density
    distribution~(see figure~\ref{fig:100nmliposomesdensity}).}
\end{figure}

The polymer segment density distributions
(figure~\ref{fig:100nmliposomesdensity}) are from Hill's
self-consistent mean-field model~\citep{Hill:2004a}. These
calculations account for surface curvature, yielding slightly less
dense layers with $a = 100$~nm than with $a = 1750$~nm. The
electrostatic potential is much lower, and, as expected, this changes
the electrophoretic mobility, as seen by comparing the mobilities in
figures~\ref{fig:1750nmliposomesmobility}
and~\ref{fig:100nmliposomesmobility} with $a = 1750$ and 100~nm,
respectively.

\subsection{PEO adsorbed on polystyrene latices} \label{sec:peo}

As demonstrated in section~\ref{sec:interpretation}, the brush-like
structure of terminally anchored PEG permits a satisfactory
uniform-layer approximation when the density and thickness yield the
`correct' hydrodynamic coating thickness. Poly(ethylene oxide) (PEO)
homopolymer, which comprises the same monomer unit as PEG, has
molecular weights an order of magnitude higher. Adsorption produces
much more inhomogeneous layers~\citep[see][and the references
therein]{Fleer:1988,Fleer:1999b}, and because desorption is slow,
experiments can be performed under conditions where the influence of
polymer on the hydrodynamic size and electrophoretic mobility is
significant, without altering the electrolyte viscosity.

Gittings and Saville~\citep{Gittings:2000} measured the adsorbed
amounts, hydrodynamic layer thicknesses, and electrophoretic
mobilities of polystyrene latices in aqueous solution with three
molecular weights of adsorbed PEO. They used the mobilities of the
bare particles to infer the (varying) surface charge density as a
function of ionic strength.  As shown in table~\ref{tab:gittingsbare},
the (apparent) charge, inferred from the standard electrokinetic model
and measured mobilities of bare latices, increases with ionic
strength.

\begin{figure}
  \centering
  \vspace{1cm}
  \subfigure[]{\includegraphics[height=6cm]{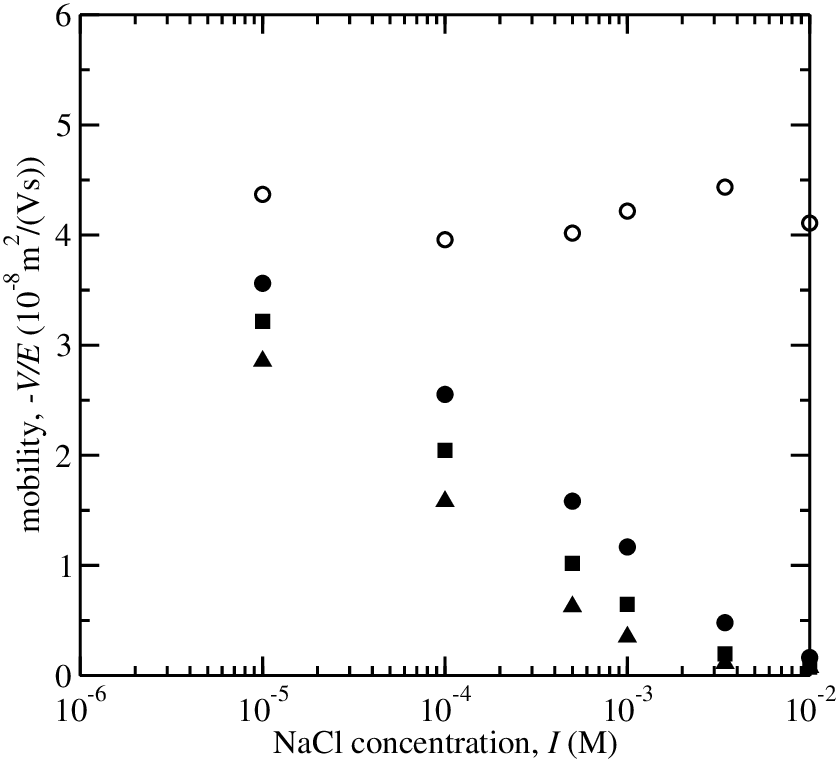}}
  \subfigure[]{\includegraphics[height=6cm]{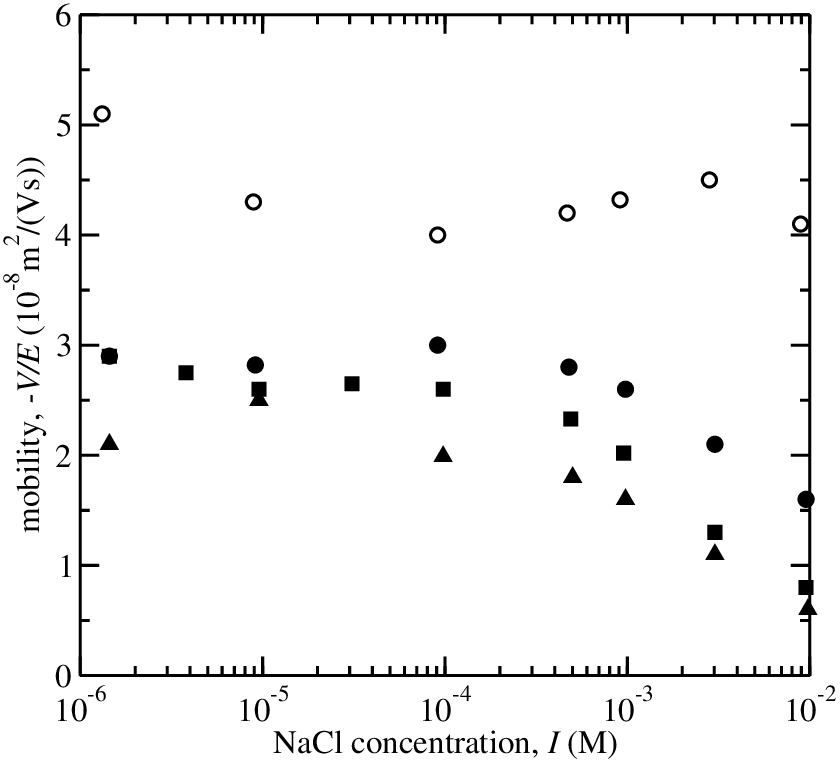}}
  \caption{\label{fig:gittingsmobility} Theoretical interpretation
    ({\em left}) of Gittings and Saville's experiments ({\em right})
    for the electrophoretic mobility $V / E$ of polystyrene latices
    with coatings of adsorbed PEO as a function of the ionic strength
    $I$ (aqueous NaCl at $T = 25$\degc) for various PEO molecular
    weights $W = 0$ (open circles), 23.5 (filled circles), 56
    (squares) and 93.75~kgmol$^{-1}$ (triangles). The surface charge
    densities and adsorbed amounts are specified according to Gittings
    and Saville~\citep{Gittings:2000} (see
    tables~\ref{tab:gittingsfuzzy} and~\ref{tab:gittingsbare}), and
    the mobilities are calculated using the full electrokinetic model
    with uniform layers yielding the measured hydrodynamic layer
    thicknesses and adsorbed amounts. The segment Stokes radius $a_s =
    0.175$~\AA \ with $l = 0.71$~nm is from Hill's
    interpretation~\citep{Hill:2004a} of Cohen and Khorosheva's
    experiments~\citep{Cohen:2001} with terminally anchored PEG on
    multilamellar liposomes.}
\end{figure}

\begin{figure}
  \centering
  \vspace{1cm}
  \subfigure[]{\includegraphics[height=6cm]{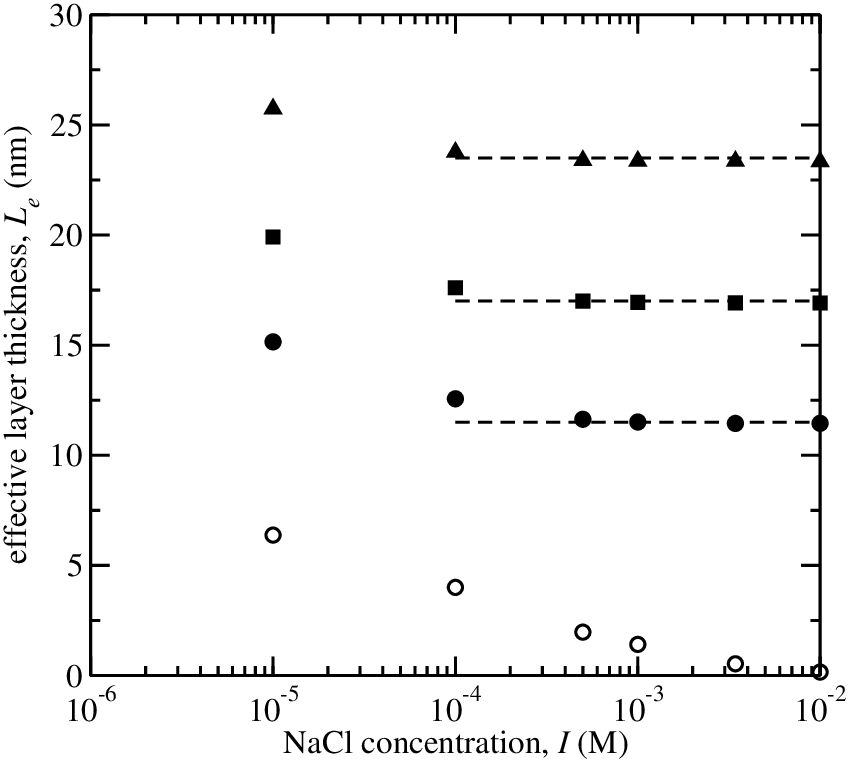}}
  \subfigure[]{\includegraphics[height=6cm]{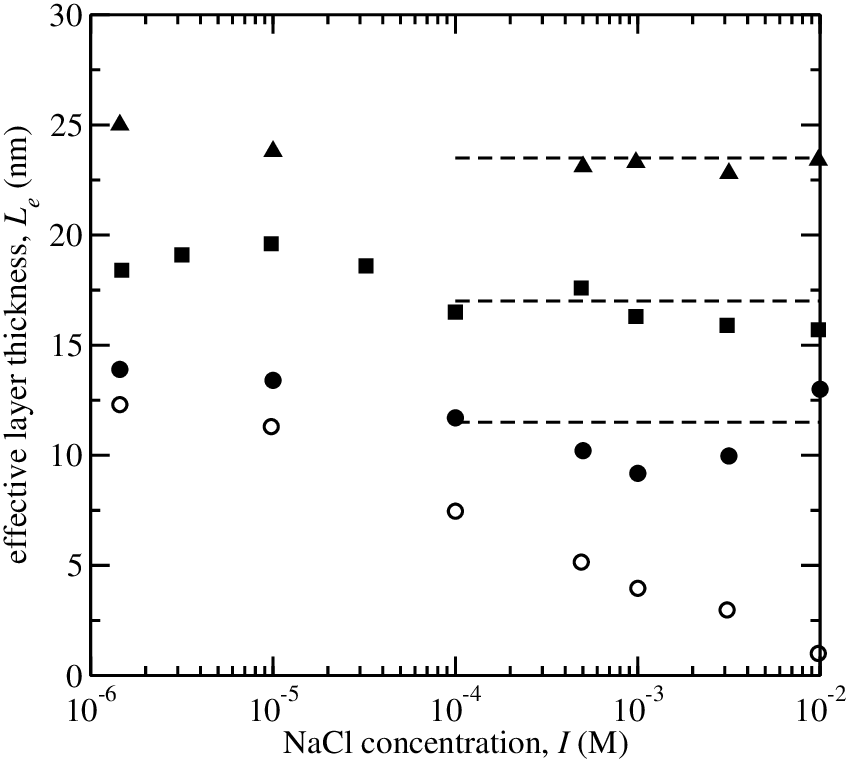}}
  \caption{\label{fig:gittingsthickness} Theoretical interpretation
    ({\em left}) of Gittings and Saville's experiments ({\em right})
    for the effective layer thickness $L_e$ of PEO adsorbed on
    polystyrene latices as a function of the ionic strength $I$
    (aqueous NaCl at $T = 25$\degc) for various PEO molecular weights
    $W = 0$ (open circles), 23.5 (filled circles), 56 (squares) and
    93.75~kgmol$^{-1}$ (triangles). The surface charge densities and
    adsorbed amounts are specified according to Gittings and
    Saville~\citep{Gittings:2000} (see tables~\ref{tab:gittingsfuzzy}
    and~\ref{tab:gittingsbare}), and the effective layer thicknesses
    are calculated using the full electrokinetic model with uniform
    layers yielding the (nominal) measured hydrodynamic layer
    thicknesses and adsorbed amounts. The segment Stokes radius $a_s =
    0.175$~\AA \ with $l = 0.71$~nm is from Hill's
    interpretation~\citep{Hill:2004a} of Cohen and Khorosheva's
    experiments~\citep{Cohen:2001} with terminally anchored PEG on
    multilamellar liposomes. Lines indicate the {\em hydrodynamic}
    layer thicknesses $L_h$.}
\end{figure}

With the methodology presented in section~\ref{sec:interpretation},
these very inhomogeneous structures are approximated as uniform
layers. Also, following Gittings and Saville, this analysis assumes
that the polymer does not influence the surface
charge. Table~\ref{tab:gittingsfuzzy} summarizes the measured and
inferred layer characteristics, and the electrophoretic mobilities and
effective coating thicknesses are shown in
figures~\ref{fig:gittingsmobility}
and~\ref{fig:gittingsthickness}.

Because adsorbed PEO produces layers with a very dense inner region
and a much more permeable periphery, approximating the layers as
uniform is crude. Nevertheless, varying the segment density
distributions to improve the predicted mobilities is invariably met
with a poorer prediction of the measured effective layer
thicknesses. In particular, adopting a model with two step-like
regions (a very dense and relatively thin inner layer with a much
thicker and permeable outer layer) produced results very similar to
those in figures~\ref{fig:gittingsmobility}(a)
and~\ref{fig:gittingsthickness}(a) with a single region. This was also
the case with exponentially decaying distributions. Varying the Stokes
radius of the segments, while maintaining the `correct' hydrodynamic
layer thicknesses and adsorbed amounts, slightly improved the
predicted electrophoretic mobilities, but the resulting layer
thicknesses $L$ were unrealistically large, and the resulting Stokes
radius $a_s$ was extremely small.

Comparing the model predictions in
figure~\ref{fig:gittingsmobility}(a) with Gittings and Saville's
measurements in figure~\ref{fig:gittingsmobility}(b) reveals that the
influence of polymer on the electrophoretic mobilities is
underestimated by the model at low ionic strength, and overestimated
at high ionic strength. The qualitative effects of varying the PEO
molecular weight and adsorbed amount are captured relatively well,
however. Comparing variations in the effective layer thicknesses of
the bare and coated particles in figure~\ref{fig:gittingsthickness}
demonstrates that polymer attenuates the electro-viscous contribution
to the drag. Even for the bare latices, variations in the effective
size with ionic strength are greater than suggested by the
theory. Recall, $L_e$ reflects an increase in the apparent radius of
the bare particle. Therefore, for colloids without a polymer coating,
$L_e$ equals the hydrodynamic thickness $L_h$ only in the absence of
electro-viscous drag.

Evidently, the influence of adsorbed polymer on the surface charge
depends on the ionic strength. In general, therefore, neutral polymer
tends to reduce the (apparent) charge more than expected from the
effect on hydrodynamic transport processes alone. Either the polymer
changes the underlying charge or charge affects the polymer
configuration. At very low ionic strength, a considerable portion of
the diffuse double layer resides outside a uniform polymer layer,
suggesting that either the polymer extends further from the interface,
to increase the effective layer thickness, or the charge is
diminished.

Note that polymer lowers the effective dielectric constant at the
interface, thereby increasing correlations between the fixed and
mobile charge, which, in turn, could decrease the degree of counterion
dissociation. Note also that polarization of low-dielectric neutral
polymer segments by the equilibrium electrostatic potential may
increase layer thicknesses, particularly at low ionic strength when
the double layer is thick and the surface potential is high. We have
not explored the possibility that the Stokes radius of the segments
may vary with the local polymer configuration. With large fluctuations
in segment density in the periphery, the effective Stokes radius might
be very small, whereas the thin dense region at the surface may
produce an effective Stokes radius more representative of the segment
size. The model also neglects the influence that dense polymer has on
molecular diffusion and electro-migration. Therefore, while the full
model improves upon earlier interpretations based on the standard
model~\citep[see][]{Gittings:2000}, further investigation is required.

  \begin{table}
    \begin{center}
      \caption{\label{tab:gittingsbare} The ionic strength-charge
	relation for the PEO-coated polystyrene latices ($a = 78$~nm)
	used in the Gittings-Saville
	experiments~\citep{Gittings:2000}.  The surface charge
	densities are inferred from the measured electrophoretic
	mobilities of the bare latices. See
	table~\ref{tab:gittingsfuzzy} for the PEO-coating parameters.}
      \begin{tabular*}{\columnwidth}{@{\extracolsep{\fill}}cc} \hline
	$I$~(M) & $\sigma_c$~($\mu$Ccm$^{-2}$) \\ \hline $10^{-5}$ &
	0.250 \\ $10^{-4}$ & 0.600 \\ $5 \times 10^{-4}$ & 0.995\\
	$10^{-3}$ & 1.352\\ $3.43 \times 10^{-3}$ & 1.713\\ $10^{-2}$
	& 1.928\\ \hline
      \end{tabular*}
    \end{center}
  \end{table}   

  \begin{table}
    \begin{center}
      \caption{\label{tab:gittingsfuzzy} Polymer layer properties for
	the PEO-coated polystyrene latices used in the
	Gittings-Saville experiments~\citep{Gittings:2000}. The
	electrophoretic mobilities of the bare and `fuzzy' latices are
	presented in figure~\ref{fig:gittingsmobility}. These
	parameters characterize the uniform layers adopted in the full
	electrokinetic model. See table~\ref{tab:gittingsbare} for the
	surface charge densities.}
      \begin{tabular*}{\columnwidth}{@{\extracolsep{\fill}}ccccccc}
	\hline $W$~(kgmol$^{-1}$) & $\sigma_p$~(mgm$^{-2}$) &
	$L_h$~$^1$~(nm) & $N$ &$L$~(nm) &
	$\bsl$~(nm) & $L_h$~$^2$~(nm)\\ \hline 
23 & 0.67 & 11.5  & 534  & 13.8 & 2.32 & 11.4 \\
56 & 0.73 & 17.0  & 1270 & 19.8 & 2.76 & 16.9\\
93.75 & 0.59 & 23.5  & 2130 & 27.3 & 3.76 & 23.3\\ \hline
      \end{tabular*}
    \end{center}
	{\small $^1$~From Gittings and Saville's
	  experiments~\citep{Gittings:2000}.\\ $^2$~Calculated with uniform layers.}
  \end{table}   

\section{The mobility of colloids with charged coatings} \label{sec:charged}

\subsection{Thin polyelectrolyte coatings} \label{sec:thinpoly}

\begin{figure}
\centering
\vspace{1cm}
\includegraphics[height=7cm]{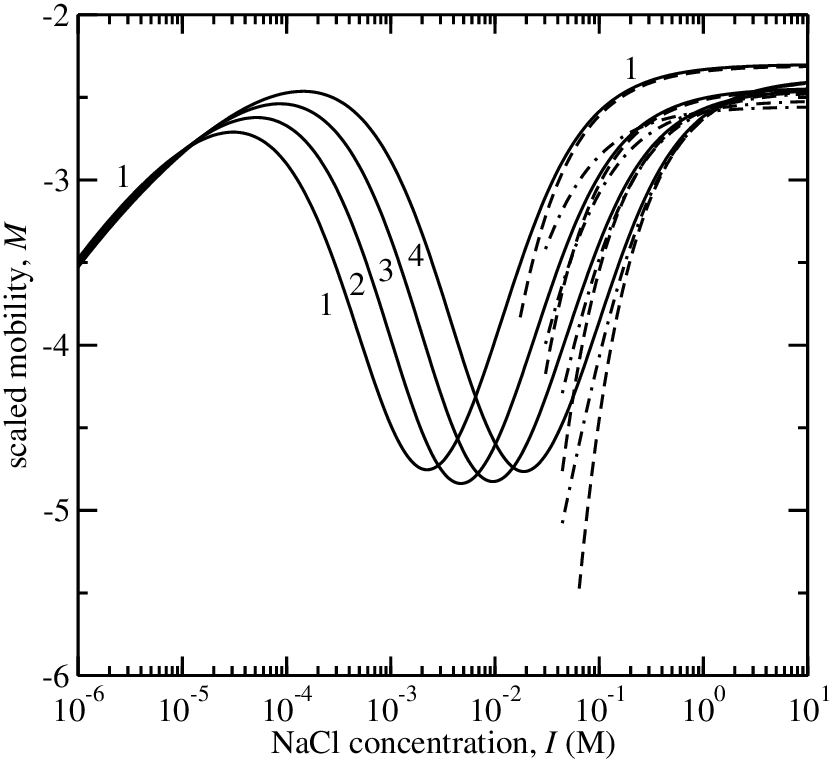}
\caption{\label{fig:largechargedmobility} The (scaled) electrophoretic
  mobility $M = 3 \eta e V / (2 \epsilon_s \epsilon_o \kb T E)$ of
  spherical colloids with relatively thin, uniform coatings of charged
  polymer as a function of the ionic strength $I$ (aqueous NaCl at $T
  = 25$\degc): $a = 500$~nm; $L = 0.25 \sqrt{10^3}$~nm; $a_s =
  0.95$~\AA; $n_0 \approx 0.132$, 0.263, 0.526 and 1.05~M (labeled
  1--4 with $\bsl \approx 2.6$, 1.8, 1.3 and 0.89~nm); and $n^f_0 =
  (2.5/ 7.1) n_0$. {\em Solid} lines are the full model with uniform
  (step-like) segment density distributions, {\em dash-dotted} lines
  are Ohshima's theory for thin uniform layers with $L / \bsl \gg
  1$~\citep[][Eqn.~(11.4.18)]{Ohshima:1995}, and {\em dashed} lines
  are Levine~\etal's theory for thin uniform layers with arbitrary $L
  / \bsl$ and $|\psi| < 2 \kb T /e$~\citep[][Eqn.~(21)]{Levine:1983}.}
\end{figure}

In this section we examine moderately large particles with thin
charged (polyelectrolyte) layers. The electrophoretic mobilities of
colloids with radius $a = 500$~nm in
figure~\ref{fig:largechargedmobility} show the effect of varying the
(uniform) polyelectrolyte density while maintaining a constant layer
thickness $L = 0.25 \sqrt{10^3}$~nm. As the ionic strength increases
from $I = 10^{-6}$ to $10$~M, $\kappa a$ spans the range 1.64--5200;
the double-layer thickness is comparable to the coating thickness when
$I \approx 1.5$~mM. The densities of the coatings span the range $n_0
\approx 0.13$--1.1~M, so with a segment Stokes radius $a_s = 0.95$~\AA
\ the respective Brinkman screening lengths span the range $\bsl
\approx 2.6$--0.89~nm. Note that Stokes radius of the (monomer)
segments adopted here is an approximate value for poly(styrene
sulfonate) monomer segments inferred by Hill~\citep{Hill:2004a} from
experiments by Cottet~\etal~\citep{Cottet:2001} reporting the
electrophoretic mobilities and hydrodynamic radii of associating
copolymer micelles~\citep{Hill:2004a} (see \S~\ref{sec:diblocks}).

The fixed charge density is assumed proportional to the segment
(monomer) density and, furthermore, is specified assuming one
elementary charge per {\em Bjerrum length} $l_B$ of the
polyelectrolyte contour length. Accordingly, the charge density $n^f_0
= (2.5/ 7.1) n_0$; each segment has length $l = 2.5$~\AA \ with $l_B =
e^2 / (4 \pi \epsilon_o \epsilon_s \kb T) \approx 7.1$~\AA. This
simple model accounts for {\em counterion condensation}, which,
according to Manning's well-known theory~\citep{Manning:1969}, limits
the {\em effective} linear charge density to $e l_B^{-1}$ when the
{\em actual} linear charge density exceeds this value. For highly
charged polyelectrolytes, \eg, poly(styrene sulfonate) with degree of
sulfonation greater than 40 percent, accounting for counterion
condensation in this way leads to `reasonable' electrostatic
potentials (Donnan potentials) inside the layer. When interpreting
measured electrophoretic mobilities, the methodology infers a Stokes
radius for the monomer segments $a_s \approx 0.95$~\AA, which, in
contrast to PEG brushes~(see~\S~\ref{sec:interpretation}
and~\S~\ref{sec:peg}), is representative of the segment (monomer)
length, $l_m \approx 2.5$~\AA.

The moderately large particle radius ($a = 500$~nm) facilitates a
comparison of the full model with analytical solutions valid as $\ka
\rightarrow \infty$ with $\kL \gg 1$. Levine~\etal's
theory~\citep[][Eqn.~(21)]{Levine:1983} ({\em dashed} lines) is in
excellent agreement with the numerically exact results when the ionic
strength is high, \ie, when $\kL \gg 1$ and the Debye-H{\"u}ckel
approximation is justified~($\psi < 2 \kb T / e$). The range of ionic
strengths over which their formula is accurate increases with
decreasing charge density, suggesting that the Debye-H{\"u}ckel
approximation is more restrictive here than the condition $\kL \gg
1$. Ohshima's theory~\citep[][Eqn.~(11.4.18)]{Ohshima:1995} ({\em
dash-dotted} lines), which neglects the viscous stress at the grafting
surface, is restricted here to dense layers with $L / \bsl > 8$.

At high ionic strength, viscous drag at the grafting surface leads to
lower mobilities than predicted by Ohshima's theory. This is
demonstrated most clearly by the particle with the least dense
layer\footnote{Close examination of
figure~\ref{fig:largechargedmobility} reveals that Ohshima's theory
({\em dash-dotted} lines) yields a mobility about 10 percent higher
(in magnitude) than the full theory at high ionic strength.}  (case 1
in figure~\ref{fig:largechargedmobility}). Under free-draining
conditions---the regime that underlies the high-ionic-strength-limit
of Ohshima's theory---the electrophoretic mobility of this particle is
expected to be the highest, since the drag coefficient of the segments
is least because of the small, but finite, (hydrodynamic) volume
fraction $\phi_s = n (4/3) \pi a_s^3 \sim 10^{-4}$. However, because
the layer is sufficiently permeable {\em and} thin, viscous stress
from the substrate also contributes to the drag. Consequently, this
particle has the lowest mobility at high ionic strength. At lower
(intermediate) ionic strength, an increasing fraction of the mobile
charge resides beyond the coating, and the mobility might therefore be
expected to increase with the total charge (increasing segment
density). However, the particles with the highest charge are more
susceptible to polarization, and this leads to an inverse relationship
between the charge and mobility. Nevertheless, at even lower ionic
strength, particles with higher charge eventually exhibit higher
mobilities. In this limit, the total charge, rather than the slightly
larger hydrodynamic radius~(friction coefficient), has the greater
influence on the mobility.

\subsection{Glycocaylx on human erythrocytes} \label{sec:bloodcells}

Levine~\etal~\citep{Levine:1983} developed their theory to interpret
the electrophoretic mobility of human erythrocytes, revealing that the
glycocaylx charge density is sufficiently high to produce an
electrostatic potential that violates the Debye-H{\"u}ckel
approximation. Sharp and Brooks~\citep{Sharp:1985} then solved the
planar electrokinetic transport problem numerically, relaxing the
Debye-H{\"u}ckel approximation, and they were able to infer a layer
thickness of $L = 7.8$~\AA, with segment and charge densities of $n_0
\approx 0.0690$ and $n_0^f \approx 0.0472$~M, respectively, at ionic
strengths in the range $I = 10$--160~mM. Note that, with a segment
Stokes radius $a_s = 7$~\AA \ and drag coefficient $F_s = 1$, the
Brinkman screening length is $\bsl \approx 1.35$~nm.  Sharp and
Brooks' calculations are limited by the finite particle size
(polarization and relaxation), so it remains to establish the range of
ionic strengths over which the flat-plate approximation is
accurate. We will also briefly examine the effect of varying the layer
thickness, while maintaining a constant mass and charge.

\begin{figure}
  \centering
  \vspace{1cm}
  \includegraphics[height=7cm]{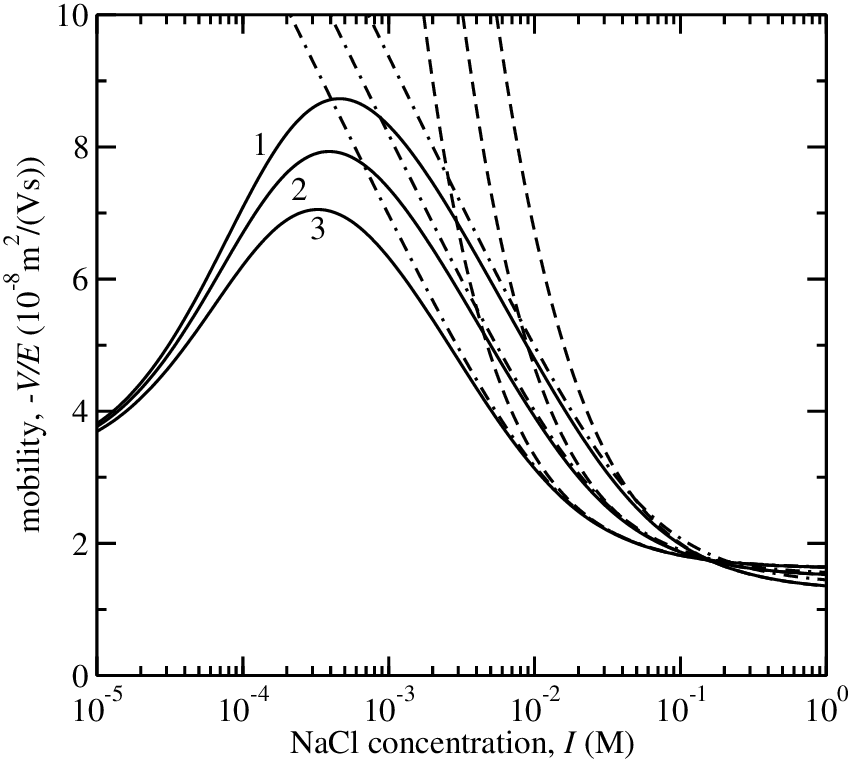}
  \caption{\label{fig:levineetalmobility} The electrophoretic mobility
    $V / E$ of human erythrocytes as a function of the ionic strength
    $I$ (aqueous NaCl at $T = 25$\degc): $a = 3.5~\mu$m; $L = 7.8/2$,
    7.8 and $7.8 \times 2$~nm (labeled 1--3, respectively); $a_s =
    3.5$~\AA; $n_0 \approx 0.0690 \times 2$, 0.0690 and $0.0690 / 2$~M
    (labeled 1--3 with $\bsl \approx 2.52$, 1.73 and 1.16~nm,
    respectively); and $n^f_0 = (0.0472/0.0690) n_0$. {\em Solid}
    lines are the full model with uniform (step-like) segment density
    distributions, {\em dash-dotted} lines are Ohshima's theory for
    thin uniform layers with $L / \bsl \gg
    1$~\citep[][Eqn.~(11.4.18)]{Ohshima:1995}, and {\em dashed} lines
    are Levine~\etal's theory for thin uniform layers with arbitrary
    $L / \bsl$ and $|\psi| < 2 \kb T
    /e$~\citep[][Eqn.~(21)]{Levine:1983}.}
\end{figure}

The mobility-ionic strength relationship in
figure~\ref{fig:levineetalmobility} verifies the accuracy of Ohshima's
theory ({\em dash-dotted} lines) for this problem when the ionic
strength is greater than approximately 5~mM. Clearly, Levine~\etal's
theory ({\em dashed} lines) is limited here by the Debye-H{\"u}ckel
approximation, even at low ionic strength. At high ionic strength,
however, a close examination reveals that their expression is more
accurate than Ohshima's, because it accounts for viscous stress on the
bare surface. Finally, the `exact' results ({\em solid} lines) show
that the particle size ($a = 3.5~\mu$m) influences the mobility when
the ionic strength is less than about 10~mM, \ie, when $\ka < 10^3$.

To assess the effect of expansion and contraction of the layers due to
electrostatic stiffening of the polyelectrolyte at low ionic strength,
we perturbed Sharp and Brooks' data (case 2) by halving (case 1) and
doubling (case 3) the (glycocaylx) layer thickness. Because the
particle radius is large ($a = 3.5~\mu$m), the changes are accompanied
by respective doubling and halving of the segment and charge
densities. Expansion decreases the segment density, thereby decreasing
the fixed charge density and increasing the layer permeability. To
account for the influence of density on the segment drag coefficient,
the Stokes radius of the segments is $a_s = 3.5$~\AA, which yields a
slightly larger Brinkman screening length ($\bsl = 1.73$~\AA) than
that of Sharp and Brooks when $L = 7.8$~\AA. The hydrodynamic volume
fraction inferred by Sharp and Brooks' parameters suggests that
varying the density should also affect the segment drag
coefficient. Indeed, we have interpreted their Stokes radius as the
product $a_s F_s(\phi_s)$, so changing the density also changes the
drag coefficient according to the change in (hydrodynamic) volume
fraction. The sensitivity of the mobility to the layer thickness and,
hence, density, is most pronounced at an ionic strength of
approximately $1$~mM. Because polarization and relaxation are
significant under these conditions, numerically exact solutions could
benefit future studies, even with relatively large particles.

\subsection{Poly(styrene sulfonate) micelles} \label{sec:diblocks}

\begin{figure}
  \centering
  \vspace{1cm}
  \includegraphics[height=7cm]{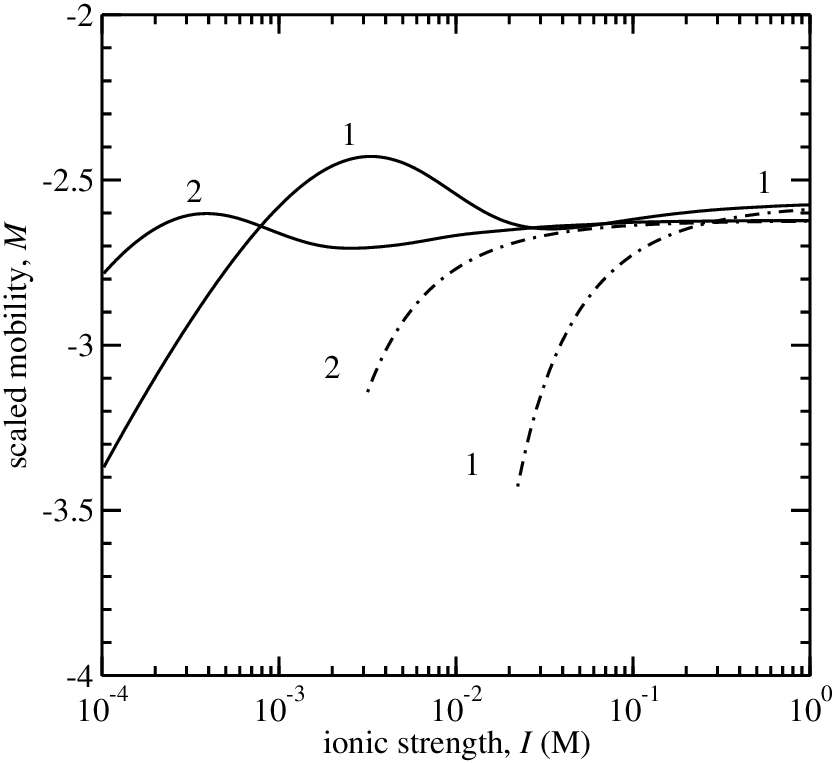}
  \caption{\label{fig:hermansfujitamobility} The (scaled)
    electrophoretic mobility $M = 3 \eta e V / (2 \epsilon_s
    \epsilon_o \kb T E)$ of spherical colloids with relatively thick,
    uniform coatings of charged polymer as a function of the ionic
    strength $I$ (aqueous NaCl at $T = 25$\degc). See
    table~\ref{tab:micelles} for parameters. For both particles, $a_s
    = 0.95$~\AA \ and $n^f_0 = (2.5/ 7.1) n_0$. {\em Solid} lines are
    the full model with uniform (step-like) segment density
    distributions, and {\em dash-dotted} lines are Ohshima's theory
    for spherical
    polyelectrolytes~\citep[][Eqn.~(11.3.11)]{Ohshima:1995}.}
\end{figure}

Finally we turn to `small' colloids with relatively thick charged
(sodium poly(styrene sulfonate))
layers. Figure~\ref{fig:hermansfujitamobility} shows the mobilities of
two representative hydrophobically associating copolymer micelles over
a range of ionic strengths. The radii of the hydrophobic cores, $a$,
and of the respective particles as a whole, $L_h + a$, are based on
neutron diffusion and light scattering measurements,
respectively~\citep[][]{Cottet:2001}.  Furthermore, the charged
coronas are assumed uniform, with {\em monomer} densities
\begin{equation}
  n_0 = \frac{n_a N_m}{(4/3) \pi a^3 [(1 + L/a)^3 - 1]},
\end{equation}
where $n_a$ is the aggregation number and $N_m$ is the number of
poly(styrene sulfonate) monomers per chain. Note that $L$ and $L_h$
are specified according to Eqn.~(\ref{eqn:constraint}); these and
other characteristics are listed in table~\ref{tab:micelles}.

The micelles assembled from the longest polyelectrolyte chains (case
2) have a much larger hydrodynamic radius $L_h + a \approx 80$~nm. The
smaller aggregation number $n_a \approx 35$ manifests in a relatively
low segment density $n_0 \approx 0.0074$~M, and, hence, lower density
of charged sites $n^f_0 \approx (2.5/7.1) n_0 \approx 0.0026$~M. Note
that, following the discussion in section~\ref{sec:thinpoly}, allowing
for counterion condensation yields an effective valence for {\em
  monomer} segments $z \approx - 2.5 / 7.1$. The total charge, which
is proportional to $z N_m n_a$, is only 30 percent lower,
however. Clearly, the lower segment density results in a much more
permeable particle, as evidenced by the large Brinkman screening
length $\bsl \approx 11.2$~nm. Following the methodology underlying
Eqn.~(\ref{eqn:constraint}), the actual `coating' thickness $L \approx
89$~nm is about 12 percent smaller than the chain contour length $l_c
\approx 404 \times 0.25 = 101$~nm, indicating that the chains are
strongly stretched.

\begin{table}
\begin{center}
  \caption{\label{tab:micelles} Parameters derived from
    data~\citep{Cottet:2001} characterizing two micelles (with charged
    coronas) whose electrophoretic mobilities are shown in
    figure~\ref{fig:hermansfujitamobility}: $a_s = 0.95$~\AA \ and
    $n^f_0 = (2.5/ 7.1) n_0$.}
  \begin{tabular*}{\columnwidth}{@{\extracolsep{\fill}}cccccc}
    \hline
    case & $a$~(nm) & $N$ & $n_a$ & $N n_a 10^{-3}$ & $L_h+a$~$^1$(nm) \\
    \hline
    1 & 5.2 & 227 & 85 & 19.3 & 42 \\
    2 & 2.6 & 404 & 35 & 14.1 & 80 \\
    \hline
\\
  \end{tabular*}
  \begin{tabular*}{\columnwidth}{@{\extracolsep{\fill}}ccccccc}
    \hline
    case & $L + a$~$^2$(nm) & $n_0$~(M) & $n^f_0$~(M) & $\bsl$~(nm) & $L / \bsl$ & $L / a$\\
    \hline
    1 & 45.4  & 0.082  & 0.029  & 3.31 & 12 & 7.7 \\
    2 & 91.2   & 0.0074 & 0.0026 & 11.2  & 7.9  & 34 \\
    \hline
  \end{tabular*}
\end{center}
{\small $^1$~Actual (measured) hydrodynamic size.\\
$^2$~From the measured hydrodynamic size and Eqn.~(\ref{eqn:constraint}).}
\end{table}   

Note that the permeabilities in table~\ref{tab:micelles} have been
established from electrophoretic mobilities reported by
Cottet~\etal~\citep{Cottet:2001} (at a single ionic strength yielding
$\kappa^{-1} \approx 2$~nm): $V / E \approx -3.4 \times 10^{-8}$ (case
1) and $-3.7 \times 10^{-8}$~m$^2$/(Vs) (case 2). From the
high-ionic-strength limit of the full electrokinetic model
\begin{equation} \label{eqn:mmobility}
  V / E = z e / (6 \pi \eta a_s F_s),
\end{equation}
where $z$ is the effective valence of a segment, these mobilities and
the `coating' parameters in table~\ref{tab:micelles} suggest a Stokes
radius for {\em monomer} segments $a_s \approx
0.95$~\AA~\citep{Hill:2004a}.

Equation~(\ref{eqn:mmobility}) is the free-draining limit of
Ohshima's~\citep[][Eqn.~(11.3.11)]{Ohshima:1995} and Hermans and
Fujita's theories for uniformly charged spherical
polyelectrolytes~\citep{Ohshima:1995}. Solutions of the full
electrokinetic model show that Eqn.~(\ref{eqn:mmobility}) also applies
to non-uniform layers when the segment and charge densities are
proportional to each other, and the coatings are thick enough for the
viscous stress on the substrate to be small compared to the frictional
drag on the polymer~\citep{Hill:2003a}.

Figure~\ref{fig:hermansfujitamobility} shows that the mobilities
predicted by the full model do not vary significantly over a wide
range of electrolyte concentrations. While Ohshima's theory ({\em
dash-dotted} lines) is in good agreement with the `exact' results at
very high ionic strength, the correction it provides to
Eqn.~(\ref{eqn:mmobility}) is very small. The analytical theory
compares more favorably with the `exact' results for the larger
particle, \ie, case 1 with larger $\kappa (a+L)$, despite the smaller
value of $L / \bsl$. Again, this points to the significance of
polarization and relaxation. Note that the high ionic strength at
which Eqn.~(\ref{eqn:mmobility}) applies usually permits polarization
and relaxation to be neglected. For these relatively small and highly
charged particles, however, the analytical theory is accurate only at
inordinately high ionic strength.

In contrast to bare particles and, indeed, particles with neutral
coatings, the mobilities reach minimum values at intermediate ionic
strengths. Here, the minima coincide with the electrostatic potential
inside the polyelectrolyte passing through $|\psi| \approx 2 \kb T /
e$. Note that the potentials would be significantly higher if we had
not accounted for counterion condensation, at least in the manner
described in section~\ref{sec:thinpoly}. With a characteristic
particle radius $a + L_h$, $\kappa (L_h + a) \approx 14$~(case 1) and
26~(case 2) when $I = 10$~mM, for example. Moreover, when $I = 1$~M,
$\kappa (L_h + a) \approx 140$ (case 1) and 260 (case 2). Once again,
with the mobilities shown in figure~\ref{fig:hermansfujitamobility},
we see that polarization and relaxation is important when the
double-layer thickness is greater than a few percent of the
characteristic particle size.

At low ionic strength, the mobility reflects a balance between the
electrical force on the fixed charge, which is proportional to $z N
n_a$, and the hydrodynamic drag, which is proportional to $a + L_h
\sim N^\nu$ $(0 < \nu < 1)$. It follows that the mobility should
increase with the aggregation number, which is indeed the case at low
ionic strength. Because the mobility is also susceptible to
polarization, and not all the counter charge resides beyond the
polymer, the permeability and particle size also play a role.

As expected, calculations (not reported here) with twice the segment
Stokes radius ($a_s = 1.9$~\AA) yield mobilities (at high ionic
strength) that are half those in
figure~\ref{fig:hermansfujitamobility}. At low ionic strength, the
mobilities are attenuated only slightly, whereas at intermediate ionic
strengths the minima cease to exist and the mobility (magnitude)
increases monotonically with decreasing ionic strength. Evidently,
decreasing the permeability decreases the effectiveness of convection
in polarizing the double layer, as expected from the inferences drawn
in section~\ref{sec:polarization} for neutral layers.

Note that the thickness and density of terminally anchored
polyelectrolytes {\em can} vary significantly with the bulk ionic
strength. Such variations have been neglected entirely. At low ionic
strength, the effective Kuhn length increases because of weakly
screened electrostatic repulsion among charged monomers along the
backbone (a so-called short-range interaction). This can lead to
layers whose hydrodynamic thickness is comparable to the polymer
contour length. Here, however, the relatively high ionic strength at
which the mobilities and sizes were reported suggests that excluded
volume (long-range) interactions are dominant because of the high
grafting density.  At lower densities, electrostatic interactions can
be much more influential~\citep[see][]{Hariharan:1998}. Moreover,
grafted chains may `collapse' at high ionic strength if the grafting
density is low, yielding much thinner, dense layers. The varying
segment and charge density that accompany such structural changes also
affect the electrostatic potential, which, in turn, influences the
ionic strength at which polarization affects the mobility. A
quantitative interpretation of these processes would benefit greatly
from experiments reporting mobilities and hydrodynamic radii over a
range of ionic strengths.

\section{Summary} \label{sec:summary}

The `full' electrokinetic model of Hill, Saville and
Russel~\citep{Hill:2003a} was applied to a variety of `soft'
colloids. Analogously to O'Brien and White's solutions of the standard
electrokinetic model, the methodology removes all approximations
imposed by earlier theories. Because of polarization and relaxation,
exact solutions of the full model were demonstrated to be as important
for `soft' colloids as the O'Brien and White methodology is for `bare'
particles.

A simple approach was described to link characteristics, such as the
polymer adsorbed amount and molecular weight, to key parameters in the
model, namely the polymer layer density, permeability and thickness,
and their radial distributions. Calculations revealed that the {\em
hydrodynamic} layer thickness, as distinguished from the actual
thickness, is the single most important layer characteristic
influencing the mobility. A convenient (and often reasonable)
approximation is to set the hydrodynamic thickness of a uniform layer
to its actual thickness minus the Brinkman screening length. It is
then straightforward (\eg, Eqn.~\ref{eqn:constraint}) to relate the
actual thickness, density and permeability (via the Stokes radius of
the segments) of a uniform coating to the grafting density (or
adsorbed amount) and polymer molecular weight for a non-uniform layer.

The full model was compared to Ohshima's analytical theories for thin
uniform, neutral layers. As expected, good agreement was found when
the double layer and coating are very thin compared to the particle
radius (less than a few percent) and polarization and relaxation are
negligible. Figures~\ref{fig:ohshimacompb} and~\ref{fig:ohshimacomp}
provide a useful reference for assessing the (coupled) influences of
layer thickness, double-layer thickness and particle size on
polarization and relaxation. Contrary to expectations, flat-plate
approximations breakdown at very high ionic strength. However, because
the mobility under these conditions is small, this unexpected behavior
is revealed only on logarithmically scaled mobility axes. Neutral
polymer was identified as attenuating polarization, but polarization
and relaxation still influence the mobility of particles with radii
less than a few microns.

The full model was applied to interpret the mobilities of (stealth)
liposomes with terminally anchored poly(ethylene glycol)
(PEG). Approximating the layers as uniform, with the `correct'
hydrodynamic layer thickness, provided a close correspondence with
experiments~\citep{Cohen:2001} {\em and} calculations that invoke much
more involved computations of the polymer-layer
structure~\citep{Hill:2004a}. There remain relatively small
differences between theory and experiment when the ionic strength and
polymer grafting density are low~\citep{Hill:2004a}. Polymer anchored
to the surface of fluid-like membranes poses complications beyond the
scope of the current electrokinetic model, which assumes immobile
surface charge and polymer. Nevertheless, based on earlier attempts to
establish the Stokes radius of aqueous PEG statistical segments,
theoretical predictions of the mobilities of 100~nm radius
(therapeutic) stealth liposomes were presented with a typical grafting
density and a variety of polymer molecular weights. These calculations
clearly demonstrate the extent to which neutral polymer coatings
decrease the apparent $\zeta$-potential, even though the actual charge
remains constant.

When applied to interpret the mobilities {\em and} hydrodynamic sizes
of polystyrene latices stabilized in aqueous electrolyte by adsorbed
poly(ethylene oxide) (PEO) homopolymer, the theory captured only the
qualitative aspects of the available
data~\citep{Gittings:2000}. Moreover, measured mobilities at low ionic
strength are lower than suggested by the theory. Evidently, adsorption
of neutral polymer lowers the charge, or the electric field due to
charge alters the polymer conformation. While model parameters (for
the polymer layers) can be adjusted to provide a good fit to measured
electrophoretic mobilities, this is inevitably accompanied by a poorer
prediction of the (measured) hydrodynamic layer thickness.

For particles with charged layers, the full model corroborated
approximate theories in their respective domains of applicability
(\eg, thin layers, at high ionic strength, with low electrostatic
potentials). For example, we adopted parameters that characterize
human erythrocytes, as established from theoretical and experimental
studies by Brooks and coworkers. We also compared the full model with
approximate analytical solutions for particles with `thick' charged
layers. These parameters were based on measured dimensions (at one
ionic strength) of polyelectrolyte micelles (with poly(styrene
sulfonate) coronas). Extrapolating the electrophoretic mobilities to
higher and lower ionic strength (with fixed dimensions and charge)
revealed that analytical theory (after Ohshima and Hermans and Fujita)
is accurate (for these particles) only at inordinately high ionic
strength. This shortcoming reflects the small size of the micelles,
their high charge density, and, hence, the strong influence of
polarization and relaxation. Note that the calculations invoked
counterion condensation, which places an upper limit on the effective
charge and, therefore, limits the prevailing electrostatic potential.

At present, the availability of sufficiently well-characterized
electrokinetic data for soft spheres is limited, so we trust that the
model and its availability will stimulate further experimental
studies. Again, it should be noted that, while this paper focused on
the electrophoretic mobility, the electrokinetic model also provides
characteristics (\eg, complex
polarizability/conductivity~\citep{Hill:2003b,Hill:2003d} and dynamic
mobility) derived from a variety of electrokinetic experiments.

\vspace{0.5cm}

RJH acknowledges support from the Natural Sciences and Engineering
Research Council of Canada (NSERC), through grant number 204542, and
the Canada Research Chairs program (Tier II).

\begin{appendix}
\end{appendix}

\bibliography{../../../bibliographies/global}

\begin{thebibliography}{10}
\expandafter\ifx\csname url\endcsname\relax
  \def\url#1{\texttt{#1}}\fi
\expandafter\ifx\csname urlprefix\endcsname\relax\def\urlprefix{URL }\fi

\bibitem{Anderson:1996}
J.~L. Anderson, Y.~Solomentsev, Hydrodynamic effects of surface layers on
  colloidal particles, Chem. Eng. Comm. 148--150 (1996) 291--314.

\bibitem{Dukhin:1970}
S.~S. Dukhin, N.~M. Semenikhin, Theory of double layer polarization and its
  effect on the electrokinetic and electrooptical phenomena and the dielectric
  permeability of dispersed systems, Kolloidn. Zh. 32 (1970) 360.

\bibitem{Rosen:1993}
L.~A. Rosen, J.~C. Baygents, D.~A. Saville, The interpretation of dielectric
  response measurements on colloidal dispersions using the dynamic stern layer
  model, J. Chem. Phys. 98~(5) (1993) 4183--4194.

\bibitem{OBrien:1978}
R.~W. O'Brien, L.~R. White, Electrophoretic mobility of a spherical colloidal
  particle, J. Chem. Soc., Faraday Trans. II 74 (1978) 1607--1626.

\bibitem{Levine:1983}
S.~Levine, K.~Levine, K.~A. Sharp, D.~E. Brooks, Theory of the electrokinetic
  behavior of human erythrocytes, Biophys. J. 42 (1983) 127--135.

\bibitem{Sharp:1985}
K.~A. Sharp, D.~E. Brooks, Calculation of the electrophoretic mobility of a
  particle bearing bound polyelectrolyte using the nonlinear
  {Poisson-Boltzmann} equation, Biophys. J. 47 (1985) 563--566.

\bibitem{Hermans:1955b}
J.~J. Hermans, H.~Fujita, Electrophoresis of charged polymer molecules with
  partial free drainage, Koninkl. Ned. Akad. Wetenschap. Proc. B58 (1955) 182.

\bibitem{Hermans:1955a}
J.~J. Hermans, Sedimentation and electrophoresis of porous spheres, J. Polym.
  Sci. 18 (1955) 527--534.

\bibitem{Ohshima:1989}
H.~Ohshima, Approximate analytical expressions for the electrophoretic mobility
  of colloidal particles with surface-charge layers, J. Colloid Interface Sci.
  130 (1989) 281--282.

\bibitem{Ho:1996}
C.~C. Ho, T.~Kondo, N.~Muramatsu, H.~Ohshima, Surface structure of natural
  rubber latex particles from electrophoretic mobility, J. Colloid Interface
  Sci. 178 (1996) 442--445.

\bibitem{Saville:2000}
D.~A. Saville, Electrokinetic properties of fuzzy colloidal particles, J.
  Colloid Interface Sci. 222 (2000) 137--145.

\bibitem{Hill:2003a}
R.~J. Hill, D.~A. Saville, W.~B. Russel, Electrophoresis of spherical
  polymer-coated colloidal particles, J. Colloid Interface Sci. 258 (2003)
  56--74.

\bibitem{Hill:2004a}
R.~J. Hill, Hydrodynamics and electrokinetics of spherical liposomes with
  coatings of terminally anchored poly(ethylene glycol): Numerically exact
  electrokinetics with self-consistent mean-field polymer, Phys. Rev. E 70
  (2004) 051046.

\bibitem{Koch:1999}
D.~L. Koch, A.~S. Sangani, Particle pressure and marginal stability limits for
  a homogeneous monodisperse gas fluidized bed: {Kinetic} theory and numerical
  simulations, J. Fluid Mech. 400 (1999) 229--263.

\bibitem{Kim:1985}
S.~Kim, W.~B. Russel, Modelling of porous media by renormalization of the
  {Stokes} equations, J. Fluid Mech. 154 (1985) 269--286.

\bibitem{Howells:1998}
I.~D. Howells, Drag on fixed beds of fibres in slow flow, J. Fluid Mech. 355
  (1998) 163--192.

\bibitem{Jackson:1986}
G.~W. Jackson, D.~F. James, The permeability of fibrous porous-media, Can. J.
  Chem. Eng. 64~(3) (1986) 364--374.

\bibitem{Biesheuvel:2004}
P.~M. Biesheuval, Ionizable polyelectrolyte brushes: brush height and
  electrostatic interaction, J. Colloid Interface Sci. 275~(1) (2004) 97--106.

\bibitem{Ohshima:1995}
H.~Ohshima, Electrophoresis of soft particles, Advances in Colloid and
  Interface Science 62 (1995) 189--235.

\bibitem{Cohen:2001}
J.~A. Cohen, V.~A. Khorosheva, Electrokinetic measurement of hydrodynamic
  properties of grafted polymer layers on liposome surfaces, Colloids and
  Surfaces A: Physiochemical and Engineering Aspects 195 (2001) 113--127.

\bibitem{Fleer:1988}
G.~J. Fleer, M.~A. Cohen~Stuart, J.~M. H.~M. Scheutjens, T.~Cosgrove,
  B.~Vincent, Polymers at Interfaces, Chapman and Hall, 1988.

\bibitem{Fleer:1999b}
G.~J. Fleer, J.~van Male, Analytical aproximations to the {Scheutjens-Fleer}
  theory for polymer adsorption from dilute solution. 2. {Adsorbed} amounts and
  structure of the adsorbed layer, Macromolecules 32~(3) (1999) 845--862.

\bibitem{Gittings:2000}
M.~R. Gittings, D.~A. Saville, The electrokinetic behavior of bare and
  polymer-coated latices, Langmuir 16 (2000) 6416--6421.

\bibitem{Cottet:2001}
P.~Cottet, H.~Gareil, P.~Guenoun, F.~Muller, M.~Delsanti, P.~Lixon, J.~W. Mays,
  J.~Yang, Capillary electrophoresis of associative diblock copolymers, J.
  Chromatography A 939 (2001) 109--121.

\bibitem{Manning:1969}
G.~S. Manning, Limiting laws and counterion condensation in polyelectrolyte
  solutions i. colligative properties, J. Chem. Phys. 51~(3) (1969) 924--933.

\bibitem{Hariharan:1998}
R.~Hariharan, C.~Biver, J.~Mays, W.~B. Russel, Ionic strength and curvature
  effects in flat and highly curved polyelectrolyte brushes, Macromolecules 31
  (1998) 7506--7513.

\bibitem{Hill:2003b}
R.~J. Hill, D.~A. Saville, W.~B. Russel, Polarizability and complex
  conductivity of dilute suspensions of spherical colloidal particles with
  uncharged (neutral) polymer coatings, J. Colloid Interface Sci. 268 (2003)
  230--245.

\bibitem{Hill:2003d}
R.~J. Hill, D.~A. Saville, W.~B. Russel, Polarizability and complex
  conductivity of dilute suspensions of spherical colloidal particles with
  charged (polyelectrolyte) coatings, J. Colloid Interface Sci. 263~(2) (2003)
  478--497.

\end{thebibliography}
\bibliographystyle{elsart-num}

\end{document}